\documentclass[preprint,12pt,3p]{elsarticle}

\usepackage{amssymb}
\usepackage{amsmath,amsfonts,amssymb}
\usepackage{graphicx}
\usepackage[caption=false]{subfig}
\usepackage{lipsum} 
\usepackage{pgfplots}
\usepackage{eurosym}
\usepackage{url}
\usepackage{multirow}
\usepackage{booktabs}
\usepackage{hhline}
\usepackage{interval}
\usepackage{epstopdf}
\usepackage{caption}
\usepackage{commath}
\usepackage[section]{placeins}
\usepackage[colorlinks]{hyperref}

\usepackage{lineno}

 \biboptions{comma,square,sort&compress}

\begin{document}

\begin{frontmatter}

\title{Identification of Behavioural Models for Railway Turnouts Monitoring}

\author[label1]{Pegah Barkhordari\corref{cor1}}
\address[label1]{DTU Electrical Engineering, Technical University of Denmark}

\cortext[cor1]{I am corresponding author}

\ead{pebark@elektro.dtu.dk}

\author[label1]{Roberto Galeazzi}
\ead{rg@elektro.dtu.dk}

\author[label2]{Alejandro de Miguel Tejada}
\address[label2]{DTU Mechanical Engineering, Technical University of Denmark}
\ead{almite@mek.dtu.dk}

\author[label2]{Ilmar F. Santos}
\ead{ifs@mek.dtu.dk}

\begin{abstract}
This study introduces a low-complexity behavioural model to describe the dynamic response of railway turnouts due to the ballast and railpad components. The behavioural model should serve as the basis for the future development of a supervisory system for the continuous monitoring of turnouts. A fourth order linear model is proposed based on spectral analysis of measured rail vertical accelerations gathered during a receptance test and it is then identified at several sections of the turnout applying the Eigensystem Realization Algorithm. The predictviness and robustness of the behavioural models have been assessed on a large data set of train passages differing for train type, speed and loading condition. Last, the need for a novel modeling method is argued in relation to high-fidelity mechanistic models widely used in the railway engineering community.       
\end{abstract}


\end{frontmatter}


\section{Introduction}
\begin{figure}[tbp]
	\centering
	\includegraphics[width=0.8\linewidth]{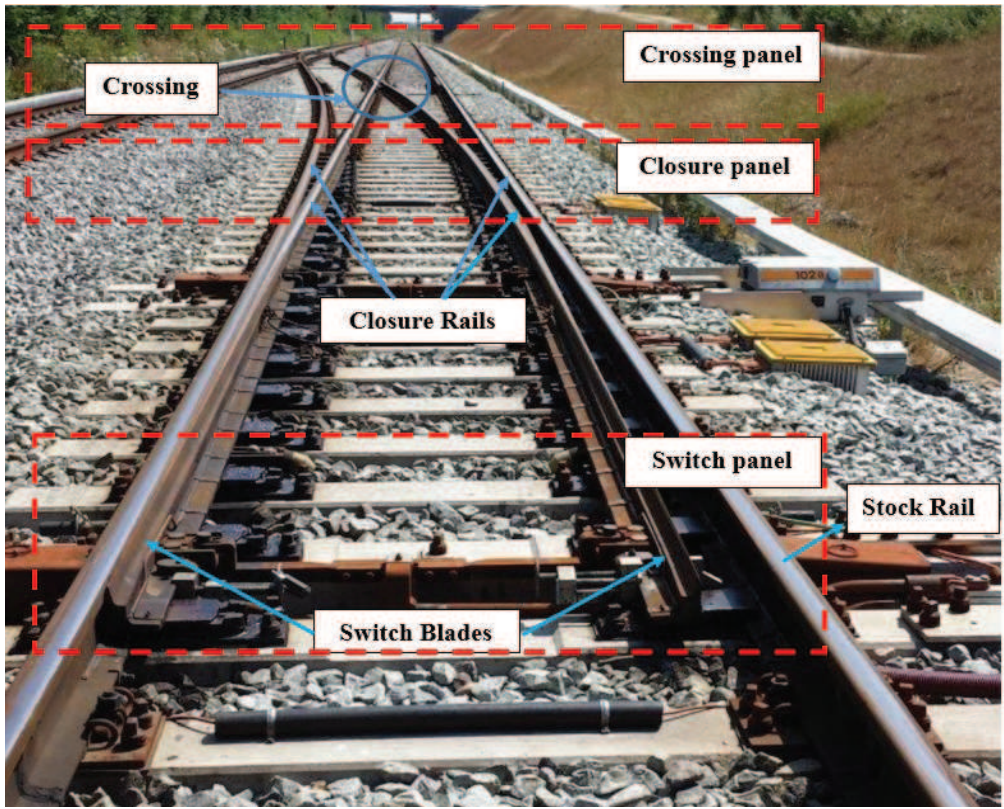}
	\caption{View of a railway turnout with highlighted sections (courtesy of Banedanmark).}
	\label{fig:SandC}
\end{figure}

\begin{figure}[tbp]
	\centering
	\includegraphics[width=0.7\linewidth]{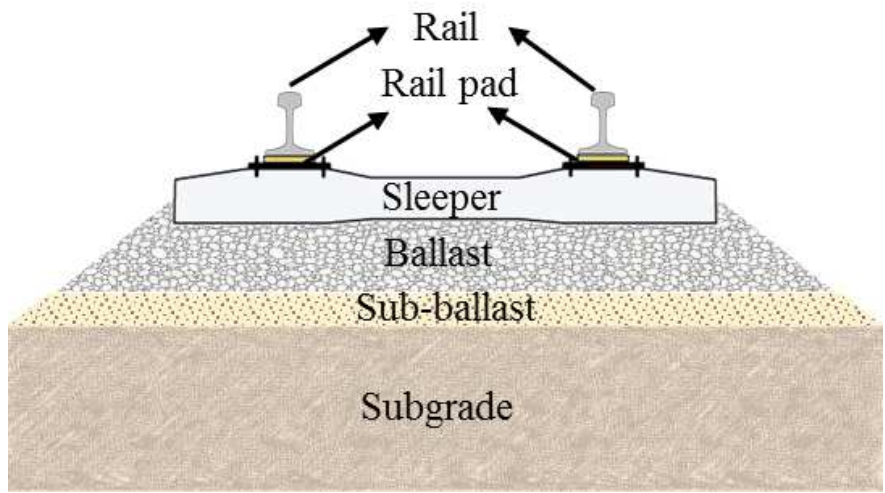}
	\caption{Railway track cross-section with highlighted components.}
	\label{fig:tracklayer}
\end{figure}

In railway networks switches and crossings (S\&C, turnout) support the trains diverging from one track to another at an angle maximizing the infrastructure utilization. As shown in Fig.~\ref{fig:SandC}, a turnout is a complex multi-component system that consists of three main consecutive sections: the switch panel, the closure panel and the crossing panel. The dynamical behaviour of a turnout at these secions results from the mechanical interaction of different track components schematically shown in Fig.~\ref{fig:tracklayer}.
The turnout geometry presents mechanical discontinuities (e.g. between closure rails and switch rails), which cause extreme impact loads on the track components upon train passage \cite{remennikov2008review,wu2003impact,hiensch2017switch}. This leads to faster degradation of the track components in the turnouts in comparison with the open track, resulting in the need of more frequent maintenance actions. Therefore, maintenance and renewal of turnouts accounts for a large share of the overall operating expenses of the infrastructure~\cite{bane2012,cornish2014life,nissen2009classification}.

If maintenance cost is to be reduced, then focus should be placed on monitoring of railway turnouts. Condition monitoring of the hidden parts of the turnouts (e.g., ballast, railpad) is the clear challenge \cite{lam2012feasibility,lam2017identification} and adoption of nondestructive methods appears the natural preferred way for infrastructure managers. Paramount for the proper operation of condition monitoring system is availability of a model which can reliably describe the dynamic behavior of the system to accurately accommodate changes in the components due to the deterioration processes. Exploiting receptance test data, this paper develops a novel model suitable for the purpose of monitoring the track components at different locations along a railway turnout.

\subsection{Literature review}
This literature review analyzes current methods for condition monitoring of the ballast layer in railway networks, discussing their advantages and disadvantages, and identifying the technological gap where the paper has the ambition to contribute. In general, the different techniques employed for condition monitoring of the railway infrastructure can be divided into two categories: \emph{direct} and \emph{indirect} methods. 

Using direct methods, the characteristics of the substructure are evaluated through explicit measurement of the quantities of interest. Among them, those most widely adopted by infrastructure managers are the ground penetrating radar (GPR)~\cite{smekal2006monitoring,berggren2009railway,kind2011gpr}, the cone penetration test (CPT)~\cite {brough2003evaluation} and visual inspection of the track at the superstructure level~\cite{labarile2004ballast,yella2009condition,asplund2013inspection}. Although the direct techniques may help infrastructure managers to schedule and perform maintenance tasks more effectively, they suffer from significant drawbacks. Difficulties in properly locating the ballast damage as well as selecting a suitable frequency range for the electromagnetic waves are known as essential restrictions of GPR. CPT is a destructive method, which may require track possession with consequent partial or complete interruption of normal train traffic. Last, the visual inspection is only useful for detection of surface damages. 

Indirect methods are non-destructive and rely on smart processing of quantities measured at the railway superstructure level (rail and sleepers). Indirect monitoring of railway infrastructure by means of measurement vehicles continuously moving along the railway track has been carried out in the past. Calculation of track stiffness, as an index of railway track quality, has been addressed in few research papers: \citet{Hosseingholian2009continuous} used a vibrating rolling wheel to excite the track and obtain the track stiffness through processing of the measured wheel acceleration; \citet{berggren2014track} used a track recording car to perform independent measurements of longitudinal track level and estimate the displacement and stiffness of the track under the wheel load.

In recent years a growing interest has emerged in developing model-based techniques, which are based on the availability of a model capable of predicting the track dynamic behavior. Different model-based techniques for ballast damage detection have been proposed. The feasibility of ballast damage detection by employing a vibration-based method based on the combination of a Timoshenko beam model and a model updating technique was studied in \cite{lam2012feasibility}. Changes in vibration characteristics (resonance frequencies and mode shapes) of the rail-sleeper-ballast model caused by reduction of sleeper support stiffness were examined and the location of the ballast damage was detected by applying an inverse method based on a time-domain model updating strategy. The suitability and robustness of the proposed framework were further investigated in~\citet{lam2014bayesian} by using a statistical description of the model unknown parameters and using a Bayesian updating technique, which explicitly accounts for model uncertainty. This work extended further in~\cite{lam2017identification} by adopting a Markov Chain Monte Carlo (MCMC)-based Bayesian model updating strategy to investigate the possibility of ballast damage detection for a system with high level of uncertainty. Results were validated using acceleration data measured in a field-test.

Experimental characterization of the dynamic behaviour of the different track components can be achieved by performing a non-destructive impact test, also known as receptance test. The analysis of the receptance function gives insight into the dynamic properties of the track by pinpointing the main resonant frequencies~\cite{man2002survey,kaewunruen2007field}. Furthermore, the data gathered during the receptance test can be used to calibrate sophisticated numerical models of the track by setting up the stiffness and damping values of the different components~\cite{costa2012track,verbraken2013benchmark}. Results from receptance test can also be used to detect defects on the rail surface~\cite{oregui2015identification} or to analyze the effect of substructure changes in the lower frequency content of the receptance function~\cite{arlaud2016receptance}.

Regarding theoretical works, the receptance curve concepts was applied to validate the dynamic behaviour of simplified 2D finite element (FEM) models against more complex 3D FEM models~\cite{nguyen2014comparison}. Oostermeijer \cite{oostermeijer2000dynamic} developed a numerical model where the receptance test was used to characterize the dynamic behaviour of a regular ballast track and an embedded rail construction. A theoretical investigation carried out by \cite{wu2000theoretical} presented the receptance function as a suitable tool to evaluate rail and wheel flexibilities.

In the context of experimental characterization of railway track several authors used receptance tests for different purposes. \citet{oregui2015experimental} applied the impact test to identify damage on the rails in the vicinity of insulated joints. \citet{kaewunruen2007field} utilized the experimentally derived receptance function to evaluate the integrity of the track structure. In most of the scientific works that deal with track receptance, both experimental and theoretical frequency response functions are combined with the purpose of calibrating the numerical models. In this regard it is worth highlighting the works presented by~\cite{alves2017calibration} where calibration and experimental validation of a dynamic FEM train/track model at a culvert transition zones is presented. In~\cite{kouroussis2016ground} the authors updated a FEM model by using  experimental receptance data to model the ground vibrations induced by and Inter-city/Inter-region train. \citet{knothe1998receptance} compared differences between measured receptances and simulation results focusing on foundation models. In~\cite{arlaud2016receptance} experimental and theoretical receptance data are combined to evaluate different track regions e.g. regular ballast track and transition zones. \citet{alves2015under} calibrated a FEM model with experimental results from a receptance test in order to get an accurate tool to predict the effect of undersleeper pads in transition zones at railway underpasses. Last, data from receptance test was used to identify the inputs of railway turnout FEM models in \cite{kassa2008dynamic} and \cite{li2014simulation}. 

Although previously developed techniques have advanced the state-of-the-art in railway infrastructure modeling and monitoring, they lead to noticeably complex diagnostic methods. High-dimensional mechanical models included in these techniques generally lack \emph{portability} (the models are fine-tuned to one specific turnout), \emph{scalability} (inclusion of additional components is laborious) and \emph{robustness} towards intrinsic and extrinsic variability such as changes in curvature radius, type of sleepers, distribution of the loading between straight and diverging track, travelling speed of the trains. Moreover, due to the large number of parameters, monitoring techniques based on high-fidelity models will face difficulties in accommodating wear and tear in the track infrastructure components along the turnout. The aforementioned features are essential for designing a monitoring system which can readily be used across the overall S\&Cs network. However, the literature survey points out that these aspects have been left in darkness.

\subsection{Contributions and novelty}
This paper presents the development of a low-complexity data-driven behavioral model able to describe the dominant dynamics of railway turnouts due to the ballast layer and railpads. By design this model is particularly suitable for use in condition monitoring systems because it encompasses the requested properties of predictiveness, robustness, portability and scalability.

The model is obtained by applying a novel approach to the processing of field hammer test data collected along an S\&C. The method decomposes the measured track vertical accelerations into the superposition of principal vibrations associated with key components of the infrastructure and then identifies an output connected second order discrete time model by applying the Eigensystem Realization Algorithm (ERA). Identification data sets have been obtained through receptance tests performed at different locations of a turnout in the Danish railway network, including open track (OT), switch panel (SP), closure panel (CLP), crossing panel (CP) and additional sections (AS) right after the crossing panel.

The identified models are locally valid in a neighborhood of the measurement point and consistently represent the ballast and railpad dynamics with estimated track resonance frequencies in good agreement with the literature~\cite{dahlberg2006track}. Further, the models well predict the train induced turnout dynamical response in the frequency range $[80,800]\,\mathrm{Hz}$, as shown through the extensive model validation where operational factors as train type, train speed and axle load have been explicitly considered. Last, the suitability of the proposed modeling approach for condition monitoring of the track infrastructure is discussed by comparing the turnout behavioural model with a high-complexity multibody simulation model.

\section{Experimental campaign and data analysis}\label{Exp-setup}
The receptance test campaign performed on an S\&C at Tommerup station (Fyn, Denmark -- October 2017) is presented and data collected are analyzed to determine the validity of the performed experiments. The receptance test allows the characterization of the main dynamic properties of the railway track components through a series of impact tests.

The receptance test is performed by impacting the top of the rail with an instrumented impact hammer and recording the track response by using accelerometers typically placed on the rail head. Measured forces and accelerations are then combined and analyzed in the frequency domain to identify the main resonant and antiresonant frequencies of the track. The receptance is the inverse of the track stiffness \cite{ewins1984modal}. Through its analysis several well-damped natural frequencies can be obtained. Resonances at low frequencies in the receptance curve are linked to track elements such as the subgrade and/or the ballast layer; whereas resonances at high frequencies are connected to track elements located on the top such as the rail and the railpads.

The informative level of the measured acceleration is assessed by means of the coherence function, which allows to determine the frequency range where the receptance data show significant cross-correlation between input and output.

\subsection{Receptance test at Tommerup station's turnout} \label{Receptancetest}
\begin{figure}[tbp]
	\centering
	\includegraphics[width=\linewidth]{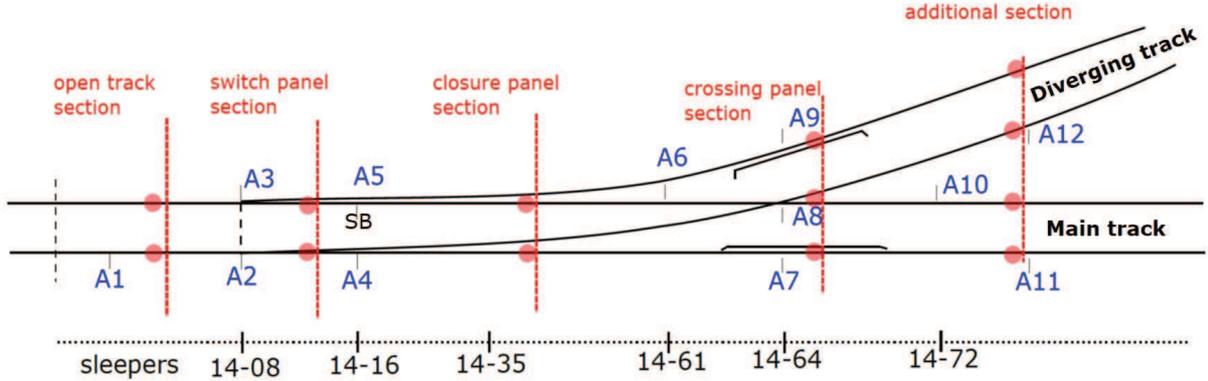}
	\caption{Layout of the measurement locations along the turnout at Tommerup station, and highlighted sections where the receptance tests have been performed.}
	\label{fig:TommerupSensorPlacement}
\end{figure}

A series of measurements was carried out at different sections along a turnout at Tommerup Station. The measurment sections with the measurement points are shown in Fig.~\ref{fig:TommerupSensorPlacement}.
The main characteristics of the instruments used to carry out the impact test are presented in Table~\ref{sensors}. An instrumented hammer with a plastic tip (086D50 PCB \cite{piezotronics2014086d50}) is used, which is suitable for exciting the track in the frequency range $[80,800]\,\mathrm{Hz}$ \cite{wei2018integrated,oregui2015experimental}. Accelerometer 1 and Accelerometer 2 are mounted on the rail head at sleeper location and  mid-span, respectively. The data collected by these two sensors during the receptance test is used to identify low-complexity behavioral models, as discussed in Section~\ref{sec:LCM}. Twelve accelerometers of type Accelerometer 3 are installed along the S\&C on the rail web and are part of a track-side measurement system employed to measure the vertical and lateral track accelerations during train passages. Measurement signals from these accelerometers are utilized to evaluate the prediction capability of the identified models.

\begin{figure}[tbp]
 \centering
	\subfloat[Graphical representation of the sensors location during the impact test.]{%
		\includegraphics[clip,width=5in,height=0.1\textheight]{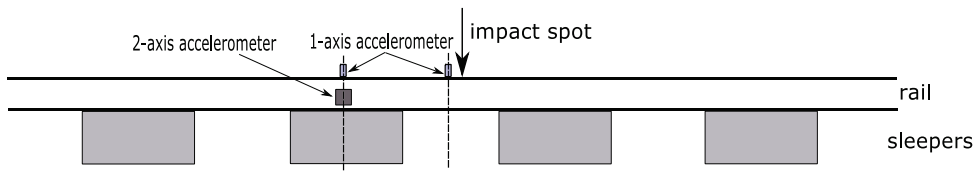}%
		\label{fig:TommerupSC}} \\
	\subfloat[Impact test performed at the additional section close to accelerometer A11.]{%
		\includegraphics[clip,width=5in]{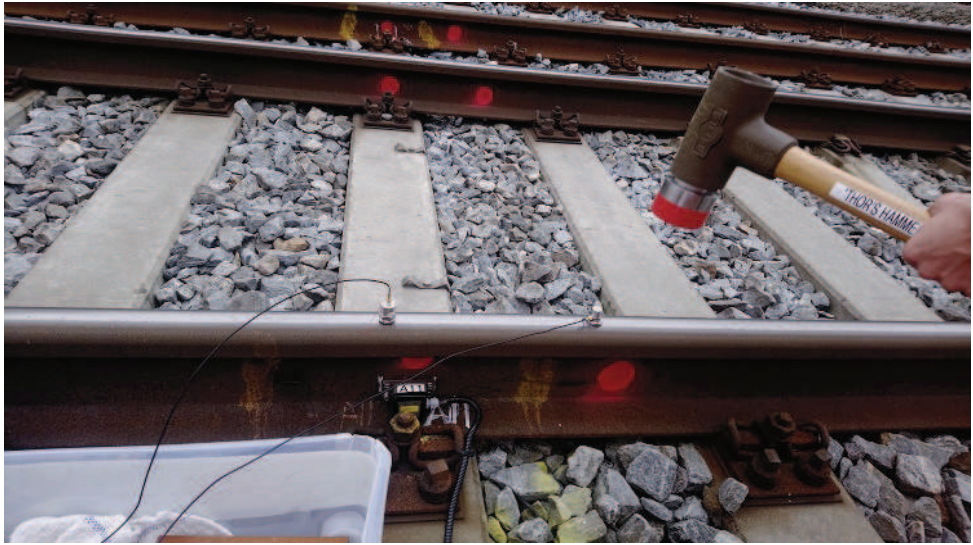}%
		\label{fig:hammering}}
	\caption{Schematic and picture of the experimental set-up employed to carry out the receptance test (Tommerup station, Fyn, Denmark - October 2017).}
	\label{fig:harmmeringtommerup}
\end{figure}

\begin{table}[tbp] \centering \small  
	\caption{Main characteristics of adopted instruments for receptance test.}\label{sensors}
	\begin{tabular}{ l*{5}{c}rl }
	\toprule
	Instrument & Brand & Range & Sensitivity & Mass \\ 
	\midrule
   	Hammer & 086D50 PCB Piezometrics &  $\pm$22.24 kN pk  & 0.23 mV/N   & 5.5 kg \\
	Accelerometer 1 & B\&K Type 4366 &  $\pm$50 g   & 35.3 mV/g   & 28 g \\
	Accelerometer 2 & B\&K Type 4370 &  $\pm$ 50 g  & 87.9 mV/g   & 54 g \\ 
	Accelerometer 3 & KISTLER 8702B500 &  $\pm$ 500 g  & 10 mV/g   & 8.2 g \\
	\bottomrule
	\end{tabular}
\end{table}
The test consisted in impacting the rail head at the impact spots shown in Fig.~\ref{fig:TommerupSensorPlacement} and measuring the vertical accelerations with the described accelerometers. Figure~\ref{fig:harmmeringtommerup} shows the set-up for the receptance test at location of A11 including the hammer and accelerometers. The schematic location of where the hammering occurs and the position of the sensors is also shown in this figure.

\begin{figure}[tbp]
	\centering
	\includegraphics[width=0.7\linewidth]{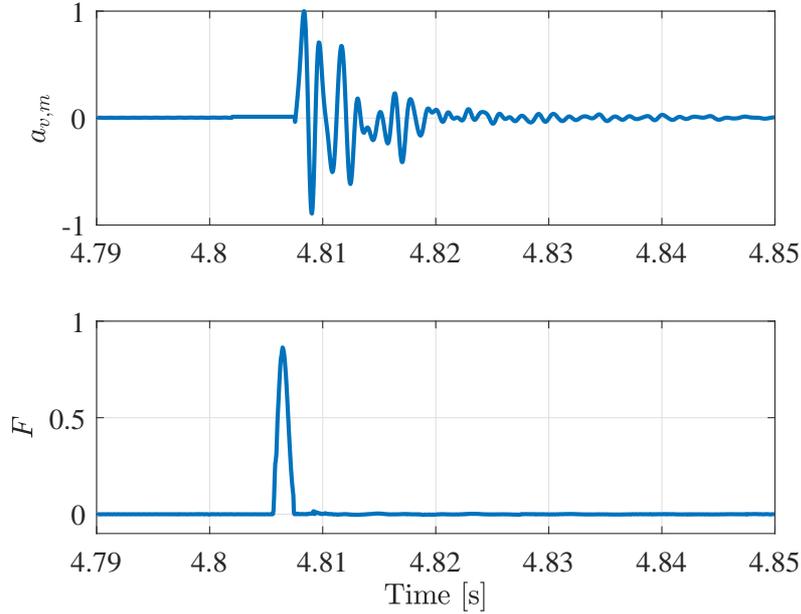}
	\caption{Outcome of the receptance test: (top) normalized vertical acceleration measured on the rail head; (bottom) measured impact force.}
	\label{fig:febtimeforce}
\end{figure}

Three receptance tests, each consisting of 10 impacts, were carried out at each measurement location. The acquisition frequency used to sample forces and accelerations was $20\,\mathrm{kHz}$. An example of the input-output data gathered during a single impact is shown in Fig.~\ref{fig:febtimeforce}. All presented measured data from the infrastructure are normalized.
\begin{figure}[tbp]
	\centering
	\includegraphics[width=0.7\linewidth]{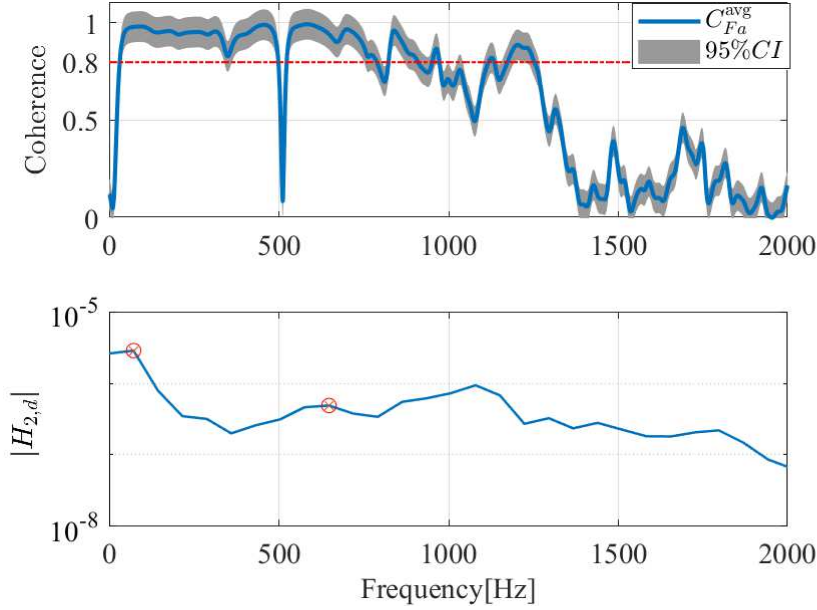}
	\caption{Analysis of receptance data: (top) average coherence function, (bottom) magnitude of $H_{2,d}$ from input-output data gathered at location A7.}
	\label{fig:coh}
\end{figure}

The quality of each experiment is evaluated computing the power transfer from hammer impact to measured vertical acceleration through the coherence function
\begin{equation}
	C_{Fa}(\omega)=\frac{\mid{G_{Fa}(\omega)}\mid^2}{G_{FF}(\omega)G_{aa}(\omega)}, \quad 0 < C_{Fa}(\omega) < 1 \label{eq:coherence}
\end{equation}
where $G_{aa}(\omega)$ refers to the power spectral density of the measured acceleration $a_{v,m}$, $G_{FF}(\omega)$ is the power spectral density of the measured force $F$, and $G_{Fa}(\omega)$ is the cross power spectral density. Under the assumption of ergodicity of the measured signals and linearity of the excited system, the coherence function $C_{Fa}(\omega)$ provides information about the range of frequencies where the system output power is produced by the input. In previous studies~\cite{oostermeijer2000dynamic,arlaud2016receptance} the coherence function was used as estimator of the quality of the signals coming from the impact test, and a $C_{Fa}(\omega) > 0.8$ was considered as strong indication of cross-correlation between the input force and the measured acceleration. We adopt the same criterion to evaluate the frequency range where the receptance tests produced valid outputs.

To account for the lack of ergodicity of the measured vertical acceleration due to non perfect repeatability of the impacts and for the presence of nonlinear effects contributing to the dynamic response of the track, the averaged coherence function $C_{Fa}^{\mathrm{avg}}(\omega)$ is calculated as the ratio of the magnitude of the averaged cross power spectral density between the measured force and the measured acceleration $G_{Fa}^{\mathrm{avg}}(\omega)$, to the product of the averaged power spectral densities of the force $G_{FF}^{\mathrm{avg}}(\omega)$ and the acceleration $G_{aa}^{\mathrm{avg}}(\omega)$ \cite{ISO1994},
\begin{equation}
C_{Fa}^{\mathrm{avg}}(\omega)=\frac{\mid{G_{Fa}^{\mathrm{avg}}(\omega)}\mid^2}{G_{FF}^{\mathrm{avg}}(\omega)G_{aa}^{\mathrm{avg}}(\omega)}, \quad 0 < C_{Fa}^{\mathrm{avg}}(\omega) < 1 \label{eq:coherence}
\end{equation}
A suitable indicator to identify resonant frequencies is given by the receptance function~\cite{ewins1984modal}
\begin{equation}
H_{2,d}(\omega)=\frac{G_{aa}(\omega)}{-\omega^2G_{Fa}(\omega)}.
\label{eq:FRF}
\end{equation}

Figure~\ref{fig:coh} shows the coherence function with the 95\% confidence interval (CI) as well as the receptance function for the receptance test performed at the measurement point A7 in crossing panel. Analyzing $C_{Fa}^{\mathrm{avg}}(\omega)$ it can be concluded that across all experiments there is a clear input-output relation up to $800\,\mathrm{Hz}$. The result shows that the coherence function is small at  $500\,\mathrm{Hz}$, indicating that the correlation between the recorded forces and accelerations is weak at this frequency. However, this is not an obstacle in the present study since the coherence function is above 0.8 at the first and second resonance frequencies. A similar behavior is observed for the receptance tests carried out at other turnout locations. Therefore, the subsequent model identification procedure only considers information content of the measured acceleration within $[0,800]\,\mathrm{Hz}$. A low-pass filter with cut-off frequency of 800 $\mathrm{Hz}$ is hence applied to the measured accelerations. Further, analyzing the magnitude of the receptance function $H_{2,d}(\omega)$ in the frequency range $[0,800]\,\mathrm{Hz}$ two significant peaks are distinguished: the first one around 100 Hz corresponds to the full track vertical resonant mode shape, whereas the second one around 600 Hz corresponds to the mode shape in which the rail bounces on the railpads. According to the literature~\cite{dahlberg2006track} these frequencies are found in the ranges of $[50,300]\,\mathrm{Hz}$ and $[200,600]\,\mathrm{Hz}$.

\section{Subspace model identification of turnout dynamics}
The Eigensystem Realization Algorithm (ERA), a subspace model identification technique, is used to identify models describing the dominant dynamic behavior of the turnout due to the ballast layer and the railpad at different sections (OT, SP, CLP, CP and~AS). 

ERA was proposed by \citet{pappa1984galileo} for model reduction using the free decay response of a system. ERA is known as an effective tool for identifying natural frequencies, mode shapes and damping ratio in structural engineering. Further, it can be combined with the impact hammer test for modal parameter identification. ERA has been successfully applied for system identification of e.g. aerospace structures~\cite{pappa1984galileo} and civil structures~\cite{caicedo2004natural}.

ERA and other methods of modal parameter identification are particularly useful in structural health monitoring. They play a significant role in identifying a model capable of predicting dominant dynamic behavior of structures. A further analysis of the dynamic characteristics of structures identified by these techniques can be carried out for damage detection and structural deterioration assessment. In civil engineering, the ERA method has been employed for structural health monitoring purposes~\cite{alvin1994second}.

The track response to a hammer excitation is a free vibration response. The system vibrates at its natural frequencies representing dynamic characteristics of the superstructure (rails and railpads) and substructure (ballast and subballast). Hence, the ERA method can be applied to the free vibration responses of the track measured during the receptance tests to identify the dynamic characteristics of the S\&C.

First, an overview of the theory behind the algorithm is provided. Development of the low-complexity behavioral models through the measured receptance test data and the ERA identification technique is then discussed in details.

\subsection{Overview of the Eigensystem Realization Algorithm}
ERA is a subspace identification method based on a state space representation of a discrete time linear time-invariant system. Let $\mathbf{x} \in \mathbb{R}^n$ be the state vector of the system to be identified, $u \in \mathbb{R}$ the known system input and $y \in \mathbb{R}$ the measured system output. Then the discrete time state-space representation of the dynamical system is
\begin{align}
    \mathbf{x}_{i+1} &= \mathbf{A} \mathbf{x}_i + \mathbf{b}u_i \label{eq:systemdynamics} \\
    y_i &= \mathbf{c}\mathbf{x}_i \label{eq:systemoutput}
\end{align}
where the subscript $i \in \mathbb{N}$ is the time index, $\mathbf{A}$ is the $n\times n$ system dynamics matrix, $\mathbf{b}$ is the $n\times 1$ input vector and $\mathbf{c}$ is the $1 \times n$ output vector. 

The free unit pulse response and the zero-input output response are given by
\begin{align}
    y_i &= \mathbf{c}\mathbf{A}^{i-1}\mathbf{b} \label{eq:pulseResponse} \\
    y_i &= \mathbf{c}\mathbf{A}^{i} \mathbf{x}_0 \label{eq:pulseResponseinitial},
\end{align}
where $\mathbf{x}_0$ is the system initial condition. Equations~\eqref{eq:pulseResponse}-\eqref{eq:pulseResponseinitial} show that at each time step the system output is given by a linear combination of the system eigenmodes. Therefore the measured output contains information enabling the identification one of the possible minimal realizations of the system $(\hat{\mathbf{A}},\hat{\mathbf{b}},\hat{\mathbf{c}})$, which is related to the true realization of the system Eqs.~\eqref{eq:systemdynamics}-\eqref{eq:systemoutput} through a similarity transformation $\mathbf{T}$, i.e. $\hat{\mathbf{A}} = \mathbf{T}^{-1}\mathbf{A}\mathbf{T}$, $\hat{\mathbf{b}}= \mathbf{T}\mathbf{b}$ and $\hat{\mathbf{c}}=\mathbf{c}\mathbf{T}^{-1}$.
    
Given the measured unit pulse response in Eq.~\eqref{eq:pulseResponse}, the Hankel matrix $\mathbf{H}_0$ and the shifted Hankel matrix $\mathbf{H}_1$ of the Markov parameters are constructed as follows~\cite{juang1985eigensystem}
\begin{equation}
\mathbf{H}_0 =
\begin{bmatrix}
{y}_1&  {y}_2  & \ldots & {y}_{n}\\
{y}_2  & {y}_3 & \ldots & {y}_{n+1}\\
\vdots & \vdots & \ddots & \vdots\\
{y}_n  &   {y}_{n+1}       &\ldots & {y}_{2n-1}
\end{bmatrix} =\begin{bmatrix}
\mathbf{cb}&  \mathbf{cAb}  & \ldots & \mathbf{cA}^{n-1}\mathbf{b}\\
\mathbf{cAb}  & \mathbf{cA}^2\mathbf{b}& \ldots & \mathbf{cA}^{n}\mathbf{b}\\
\vdots & \vdots & \ddots & \vdots\\
\mathbf{cA}^{n-1}\mathbf{b}  &   \mathbf{cA}^{n}\mathbf{b}       &\ldots &\mathbf{cA}^{2n-2}\mathbf{b}
\end{bmatrix}
\end{equation}
\begin{equation}
\mathbf{H}_1 =
\begin{bmatrix}
{y}_2&  {y}_3  & \ldots & {y}_{n+1}\\
{y}_3  & {y}_4 & \ldots & {y}_{n+2}\\
\vdots & \vdots & \ddots & \vdots\\
{y}_{n+1}  &  {y}_{n+2}       &\ldots & {y}_{2n}
\end{bmatrix}
\label{eq:shiftedhankel} 	
\end{equation}
where the dimension of the Hankel matrix is $n\times n$. The matrix $\mathbf{H}_0$ can be rewritten as
\begin{align}	
\mathbf{H}_0 &= 	\begin{bmatrix}
\mathbf{c}\\
\mathbf{cA}  \\
\vdots \\
\mathbf{cA}^{n-1} 
\end{bmatrix}
\begin{bmatrix}
\mathbf{b}	&  \mathbf{Ab}  & \ldots & \mathbf{A}^{n-1}\mathbf{b} 
\end{bmatrix}={\mathbf{\Phi}_o}{\mathbf{\Phi}_c},
\label{eq:hankel}
\end{align} 
where $\mathbf{\Phi}_o$ and $\mathbf{\Phi}_c$ are the observability and controllability matrices. Two equivalent matrices can be obtained by Singular Value Decomposition (SVD) of $\mathbf{H}_0$   
\begin{equation}
\mathbf{H}_0=\mathbf{U}\mathbf{\Sigma}^2\mathbf{V}^T=(\mathbf{U}\mathbf{\Sigma})(\mathbf{\Sigma}\mathbf{V}^T)=\mathbf{P}\mathbf{Q}. \label{eq:SVD1}
\end{equation}
Noteworthy that the matrices $\mathbf{P}$ and $\mathbf{Q}$ are not unique. According to Eqs.~\eqref{eq:shiftedhankel}-\eqref{eq:hankel} the shifted Hankel matrix can be rewritten as
\begin{align}
\mathbf{H}_1 &= \mathbf{\Phi}_o \mathbf{A} \mathbf{\Phi}_c. 
\end{align}
Hence, the system dynamics matrix $\mathbf{A}$ can be obtained from the latter equation as 
\begin{align}
\mathbf{A}=\mathbf{\Phi}_o^{-1}\mathbf{H}_1 \mathbf{\Phi}_c^{-1}.
\end{align}
Since the state space representation of the system taken into account is minimal and the system is single-input single-output, the observability and controllability matrices are full-rank and square matrices. Therefore, their invertibility is guaranteed.    

An estimate of the system dynamical matrix is obtained by utilizing the matrices $\mathbf{P}$ and $\mathbf{Q}$ equivalent to the observability and controllability 
\begin{align}
\hat{\mathbf{A}}=\mathbf{P}^{-1}\mathbf{H}_1 \mathbf{Q}^{-1}.
\end{align}
Estimates of the input and the output vectors ($\hat{\mathbf{b}}$ and $\hat{\mathbf{c}}$) are obtained by taking the first column of the matrix $\mathbf{Q}$ and the first row of the matrix $\mathbf{P}$
\begin{align}
\mathbf{P} &= \begin{bmatrix}
\mathbf{\hat{ c}}\\
\mathbf{\hat{c} \hat{A}}  \\
\vdots \\
\mathbf{ \hat{c} \hat{A}}^{n-1} 
\end{bmatrix} \, , \,
\mathbf{Q} =
\begin{bmatrix}
\mathbf{\hat{ b}}	&  \mathbf{\hat{A} \hat{b}}  & \ldots &\mathbf{\hat{A}}^{n-1} \mathbf{\hat{b}} 
\end{bmatrix}
\label{eq:matrixhankel}
\end{align}

Given the identified system dynamical matrix $\mathbf{\hat{A}}$ the modal properties of the system in terms of natural frequencies and damping ratios can be computed using the following formulas 
\begin{equation}
\omega_{nk} = \frac{\mid{\ln({\lambda}_k(\mathbf{\hat{A}}))\mid}}{2{\pi}T_s}, \quad \zeta_{k} =\frac{-\mathrm{Re}(\ln({\lambda}_k(\mathbf{\hat{A}}))/T_s)}{\mid{\ln({\lambda}_k(\mathbf{\hat{A}}))/T_s\mid}} \label{eq:freq_damp_est}
\end{equation}  
where $T_s$ is the sampling time and $\lambda_k(\hat{\mathbf{A}})$ is the $k$-th eigenvalue of the matrix $\hat{\mathbf{A}}$.

Using the identified models, the vertical accelerations at different location along the turnout (i.e. OT, SP, CLP, CP, AS) can be estimated as~\cite{majji2010time}
\begin{equation}
\mathbf{Y}=\mathbf{P}\mathbf{\hat{x}_0}
\end{equation}
where $\mathbf{\hat{x}_0}$ is a recursive estimation of the system initial condition at the moment of impact excitation. 

\subsection{Low-complexity behavioral models} \label{sec:LCM}
\begin{figure}[tbp]
	\centering
	\includegraphics[width=0.8\linewidth]{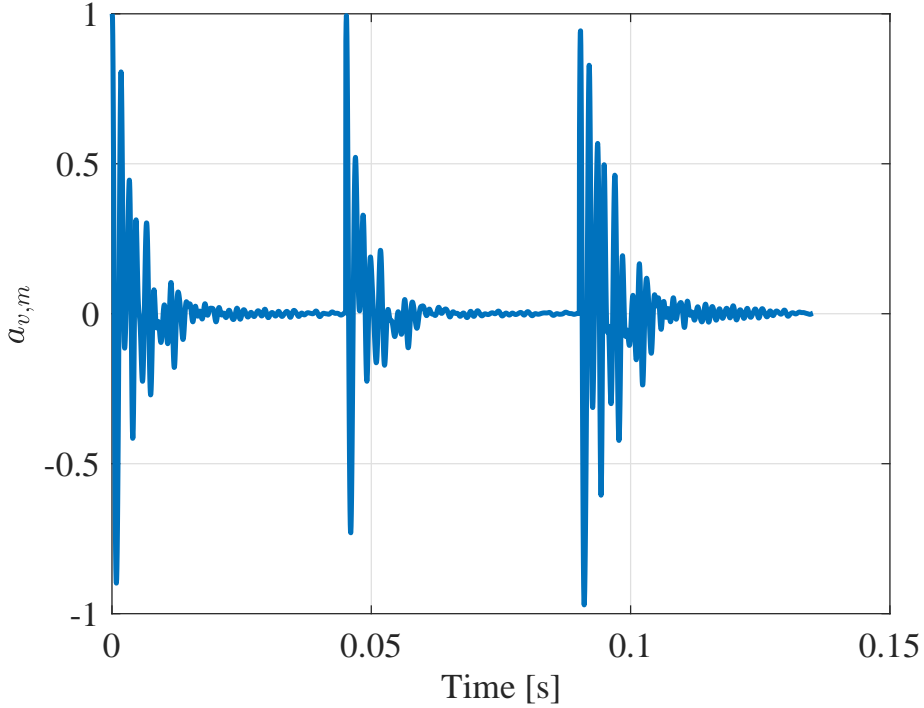}
	\caption{An example of identification data set for location A7.}
	\label{fig:identA7}
\end{figure}

To identify low-complexity behavioural models for the different locations along the S\&C identification data sets have been created by randomly selecting impact tests among those available at each location. As an example, the selected identification data set for the crossing panel section (A7) is shown in Fig.~\ref{fig:identA7}. The order of the model to be identified is determined by joint inspection of the receptance function (see Fig.~\ref{fig:coh}) and power spectral density of the vertical acceleration within the frequency range $[0,800]\,\mathrm{Hz}$. The analysis of the receptance function suggested the presence of two main resonance peaks. 

\begin{figure}[t]
	\centering
	\includegraphics[width=0.8\linewidth]{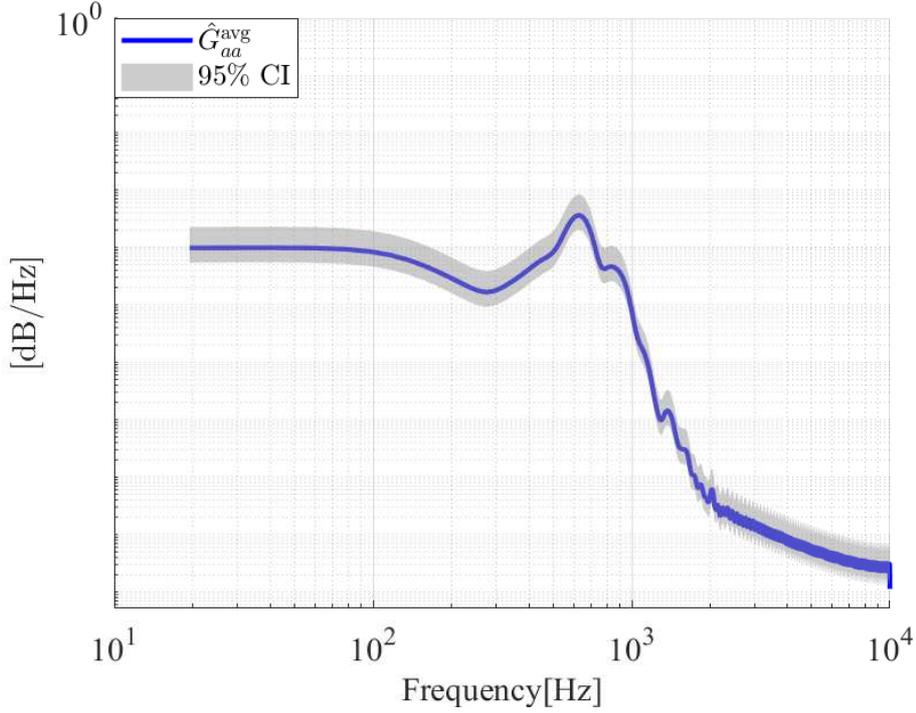}
	\caption{Average power spectrum with $95 \%$ confidence interval for vertical track accelerations measured at location A7.}
	\label{fig:psdA7}
\end{figure}

To take into account all track vertical acceleration responses measured during the three receptance tests, the average of the power spectral density over all the responses, $G_{aa}^i$, is calculated at each location. For the crossing panel (A7) the average power spectral density $G_{aa}^{\mathrm{avg}}$ of the all measured vertical accelerations with the 95\% confidence interval is illustrated in Fig.~\ref{fig:psdA7}. It is worth noting that the 95\% confidence interval is determined utilizing a chi-squared approach. When the spectral density is plotted on a logarithmic scale, the $(1-\alpha) \times 100\%$ confidence interval is constant at every frequency~\cite[Chapter 5]{manolakis2005statistical}
\begin{equation}\small
\left(10 \ln(\hat{G}_{aa}^{\mathrm{avg}}(e^{j\omega}))-10 \ln{\frac{{\chi}^2_\nu(1-{\alpha}/2)}{\nu}},\right.  
\left. 10 \ln(\hat{G}_{aa}^{\mathrm{avg}}(e^{j\omega}))+10 \ln{\frac{\nu}{{\chi}^2_\nu({\alpha}/2)}}\right)
\label{eq:confint}
\end{equation}
where $\hat{G}_{aa}^{\mathrm{avg}}$ is the average of the estimates of the power spectral density and
\begin{equation}
\nu=\frac{2N}{\sum_{l=-(L-1)}^{L} {\omega}^2_a (l)}
\end{equation}
is the degree of freedom of a ${\chi}^2_{\nu}$ distribution. $N$, $L$ and ${\omega}_a$ are the number of observations, window size and correlation window, respectively. $L$ and ${\omega}_a$ are considered to be $N$ and~$0.5N$. 

The average power spectral density $\hat{G}_{aa}^{\mathrm{avg}}$ at different locations along the turnout shows two resonance peaks in the frequency range $[0,800]\,\mathrm{Hz}$. The peak in the frequency range $[100 -200]\,\mathrm{Hz}$ is associated to the ballast layer, describing the in-phase vibrations of rail and sleeper. The out-of-phase motions of rail and sleeper occurs at the second track resonance frequency, representing the railpad effect. The analysis of the receptance function and of the measured acceleration suggests that the least complex model capturing both ballast and railpad dynamics is of order four.

Figure~\ref{fig:psdA7} shows that the amplitude of the second resonance mode is significantly larger than the amplitude of the first mode. The same trend is observed at all other locations along the turnout. Therefore, to avoid erroneous estimation of the first track resonance due to resonance peaks level discrepancy, the identification data set is split into two frequency bands using a low-pass and high-pass filter with cut-off/cut-in frequency of $200\,\mathrm{Hz}$. 

A general form of the identified models in the low ($10-200 \ \mathrm{Hz}$) and high ($200 -800 \ \mathrm{Hz}$) frequency ranges are
\begin{equation}
\mathcal{M}_{l}:\left\lbrace
\begin{aligned}
\hat{\textbf{A}}_{l} &= \begin{bmatrix} a_{l_{11}} & a_{l_{12}} \\ a_{l_{21}} & a_{l_{22}} \end{bmatrix} \\ 
\hat{\textbf{C}}_{l} &= \begin{bmatrix} c_{l_{11}} & c_{l_{12}} \end{bmatrix} \\
\end{aligned}\right.  \qquad
\mathcal{M}_{h}:\left\lbrace \begin{aligned}
\hat{\textbf{A}}_{h} &= \begin{bmatrix} a_{h_{11}} & a_{h_{12}}\\ a_{h_{21}} & a_{h_{22}} \end{bmatrix} \\
\hat{\textbf{C}}_{h} &= \begin{bmatrix} c_{h_{11}} & c_{h_{12}} \end{bmatrix} \\     
\end{aligned}\right. . \label{eq:low_highpass_ERA}
\end{equation}
The final identified model representing the dominant behavior of the vertical track dynamics due to the ballast and railpad is obtained by output connecting the two identified models
\begin{equation}
\mathcal{M}:\left\lbrace
\begin{aligned}
\hat{\mathbf{A}} &= \begin{bmatrix} \hat{\mathbf{A}}_{l} &\mathbf{0}\\ \mathbf{0}&  \hat{\mathbf{A}}_{h} \end{bmatrix} \\ 
\hat{\mathbf{C}} &= \begin{bmatrix} \hat{\mathbf{C}}_{l} & \hat{\mathbf{C}}_{h} \end{bmatrix} \\
\end{aligned}\right. . \label{eq:modelERA}
\end{equation}

The identified model in Eq.~\eqref{eq:modelERA} corresponds to the zero-input output response of the system as shown in Eq.~\eqref{eq:pulseResponseinitial}. The impact force applied in the receptance tests is available for the purpose of system identification, however no information about the wheel dynamic load is available for the measured train-induced accelerations. Hence a model based on the knowledge of the input cannot be used for validation on the train related data sets. Therefore to validate the predictiveness of the model, the initial condition is estimated for each impact (hammer or train wheel) and the overall measured response is treated as a sequence of zero-input output responses.

Tables~\ref{table1} and \ref{table2} report the estimated parameters with their uncertainties of the identified state space realization of the low and high frequency models at different locations along the turnout. The modal characteristics of the final identified models (i.e. eigenmodes, natural frequencies and damping ratios) are reported in Table~\ref{table3}. Damping ratios ($\zeta$) and natural frequencies ($\omega_n$) are calculated utilizing Eq.~\eqref{eq:freq_damp_est}. 
\begin{table}[tbp] \centering \small  
	\caption{Estimated parameters of the identified low frequency models}\label{table1}
		\resizebox{\textwidth}{!}{
	\begin{tabular}{ l*{6}{c}rl }
	\toprule
	Section &$ \hat{\mathbf{A}}_{l} $  & $\hat{\mathbf{C}}_{l}$   \\ 
	\midrule
    	OT    &  $\begin{bmatrix} 0.9988\pm 0.0004 &  -0.0481 \pm 0.0038 \\ 0.0481 \pm 0.0038 & 0.9803\pm 0.0030 \end{bmatrix}$ & $\begin{bmatrix} 0.0618 \pm 0.0018 & 1.0313\pm 0.0214 \end{bmatrix}$ \\ \\
    	
		SP    &  $\begin{bmatrix} 0.9969 \pm 0.0004 & -0.0524\pm 0.0022 \\ 0.0524\pm 0.0022  & 0.9966\pm 0.0004 \end{bmatrix}$ & $\begin{bmatrix} 0.2261\pm 0.0242 & 1.0110\pm 0.0287 \end{bmatrix}$   \\  \\
		
		CLP    &  $\begin{bmatrix} 0.9967\pm 0.0004 & -0.0326\pm 0.0069 \\ 0.0326 \pm 0.0069 & 0.9887 \pm 0.0110 \end{bmatrix}$ & $\begin{bmatrix} 0.5373 \pm 0.0108 & 0.8676\pm 0.02335 \end{bmatrix}$   \\ \\
		
		CP(A7)    &  $\begin{bmatrix} 0.9986\pm 0.0001 & -0.0307\pm 0.0016 \\ 0.0307\pm 0.0016 & 0.9824\pm 0.0011 \end{bmatrix}$ & $\begin{bmatrix} 0.3642\pm 0.0537 & 1.2523\pm 0.0226 \end{bmatrix}$   \\  \\
		
		CP(A8) &  $\begin{bmatrix} 0.9964\pm 0.0005 & -0.0542\pm 0.0009 \\ 0.0542\pm 0.0009 & 0.9899\pm 0.0014 \end{bmatrix}$ & $\begin{bmatrix} 1.1782\pm 0.0273 & 2.0409\pm 0.1328 \end{bmatrix}$   \\  \\
	
		CP (A9)   &  $\begin{bmatrix} 0.9809\pm 0.0001 & -0.0359\pm 0.0053 \\ 0.0359\pm 0.0053 & 0.9987\pm 0.0011 \end{bmatrix}$ & $\begin{bmatrix} 1.2712\pm 0.0067 & 0.1894\pm 0.0129 \end{bmatrix}$   \\  \\

		AS (main track)    &  $\begin{bmatrix} 0.9973\pm 0.0002 & -0.0482\pm 0.0075 \\ 0.0482\pm 0.0075 & 0.9830\pm 0.0032 \end{bmatrix}$ & $\begin{bmatrix} 0.3147 \pm 0.04388 & 0.8447\pm 0.0608 \end{bmatrix}$   \\ \\

        AS  (diverging track)  &  $\begin{bmatrix} 0.9961 \pm 0.0004 & 0.0406\pm 0.00820 \\ -0.0406\pm 0.0820  & 0.9888\pm 0.0036 \end{bmatrix}$ & $\begin{bmatrix} 0.5895\pm 0.0193 & 0.9803\pm 0.0940 \end{bmatrix}$   \\
        \bottomrule
        \end{tabular}}
\end{table}

\begin{table}[tbp] \centering \small  
	\caption{Estimated parameters of the identified high frequency models}\label{table2}
	\resizebox{\textwidth}{5cm}{
	\begin{tabular}{ l*{6}{c}rl }
		\toprule
		Section &$ \hat{\mathbf{A}}_{h} $  & $\hat{\mathbf{C}}_{h}$   \\ 
		\midrule
    	OT  &  $\begin{bmatrix} 0.9626 \pm 0.0002 & -0.1882 \pm 0.0005 \\ 0.1882 \pm 0.0005 & 0.9755 \pm 0.0000 \end{bmatrix}$ & $\begin{bmatrix} 2.7193\pm 0.0003 & 1.1605\pm 0.0001 \end{bmatrix}$ \\ \\

		SP    &  $\begin{bmatrix} 0.9764\pm 0.0012 & -0.1550\pm 0.0003 \\ 0.1550\pm 0.0003 & 0.9854\pm 0.0001 \end{bmatrix}$ & $\begin{bmatrix} 0.2341\pm 0.0081 & 1.0023\pm 0.0088 \end{bmatrix}$   \\ \\

		CLP    &  $\begin{bmatrix} 0.9918\pm 0.0045 & 0.1270\pm 0.0049 \\ -0.1270\pm 0.0049 & 0.9809\pm 0.0031 \end{bmatrix}$ & $\begin{bmatrix} 0.1738\pm 0.0369 & 1.7160\pm 0.0245 \end{bmatrix}$   \\  \\

		CP(A7)    &  $\begin{bmatrix} 0.9612\pm 0.0089 & -0.1898\pm 0.0015 \\ 0.1898\pm 0.0015 & 0.9768 \pm 0.0026  \end{bmatrix}$ & $\begin{bmatrix} 3.3554 \pm 0.03557 & 1.0429\pm 0.05262 \end{bmatrix}$   \\ \\
		
		CP(A9)    &  $\begin{bmatrix} 0.9763\pm 0.0006 & -0.1694\pm -0.0011 \\ 0.1694\pm 0.0011 & 0.9777\pm 0.0005 \end{bmatrix}$ & $\begin{bmatrix} 2.8610\pm 0.1067 & 2.0952\pm 0.0639 \end{bmatrix}$   \\ \\
		
		AS (main track)   &  $\begin{bmatrix} 0.9695\pm 0.0016 & -0.1955\pm 0.0013 \\ 0.1955 \pm 0.0013 & 0.9782 \pm 0.0014 \end{bmatrix}$ & $\begin{bmatrix} 2.3626 \pm 0.0781 & 0.7891\pm 0.0768 \end{bmatrix}$   \\ \\
	
		AS (diverging track)   &  $\begin{bmatrix} 0.9421\pm 0.0648 & -0.2104\pm 0.0112 \\ 0.2104 \pm 0.0112 & 0.9679 \pm 0.0201 \end{bmatrix}$ & $\begin{bmatrix} 2.8449 \pm 0.0396 & 0.9905\pm 0.0369 \end{bmatrix}$   \\
		\bottomrule
		\end{tabular}}
\end{table}

\begin{table}[tbp] \centering \small  
	\caption{Modal characteristics of the identified models.}\label{table3}
	\resizebox{!}{5cm}{
	\begin{tabular}{ l*{6}{c}rl }
	\toprule
		Section & $\lambda$ [-] & $ \omega_{n}$ [Hz]  & $\zeta$ [-]   \\ 
	\midrule
   	\multirow{2}[3]{*}{OT}  & $0.9895\pm 0.0472i$ & $154.5534 \pm 11.7249 $ & $0.1932 $  \\  &$0.9691\pm 0.1881i$ & $611.5569 \pm 1.6490$   & $0.0673 $  \\  \\

	\multirow{2}{*}{SP}  & $0.9968\pm 0.0524i$ & $167.1953 \pm 6.9777$ & $0.0353 $  \\  &$0.9814\pm 0.1553i$ & $499.1206 \pm1.0661$ & $0.0441$  \\ \\
			
	\multirow{2}{*}{CLP}  & $0.9927\pm 0.0324i$ & $105.8538 \pm 19.9734$ & $0.2042$  \\  &$0.9864\pm 0.1269i$ & $407.6153 \pm 13.4926$ & $0.0432$  \\  \\
			
    \multirow{2}{*}{CP (A7)}  & $0.9905\pm 0.0296i$ & $99.5863 \pm 5.2133$ & $0.2917 $  \\ &$0.9690\pm 0.1896i$ & $616.4479 \pm 1.0853$ & $0.0656$  \\  \\
		
	\multirow{2}{*}{CP (A8)}  & $0.9931\pm 0.0541i$ & $174.0117 \pm 2.6834 $ & $0.0986$  \\ &$-$ & $-$ & $-$  \\   \\
		
	\multirow{2}{*}{CP (A9)}  & $0.9899 \pm 0.0347i$ & $115.7699 \pm 17.6306$ & $0.2648$  \\ &$0.9770\pm 0.1694i$ & $547.6198 \pm 3.7399$ & $0.0491$  \\  \\
		
	\multirow{2}{*}{AS (main track)}  & $0.9902\pm 0.0476i$ & $155.5657 \pm 24.4137$ & $0.1788$  \\  &$0.9739\pm 0.1955i$ & $630.8860 \pm 2.9081$ & $0.0304$  \\ \\
			
	\multirow{2}{*}{AS (diverging track)}  & $0.9925\pm 0.0404i$ & $131.3086 \pm 25.3344$ & $0.1633 $  \\  &$0.9550\pm 0.2100i$ & $692.6417 \pm 5.9791$ & $0.1033$  \\ 
	\bottomrule
    \end{tabular}}
\end{table}

\subsection{Model validation and robustness}
A detailed investigation is now carried out to examine the prediction capability of the identified models. The prediction capability is tested by considering the track response to both hammer impact (using the measurement signals recorded by Accelerometer 2) and ordinary train traffic excitation measured at different locations of the turnout (using the measurement signals recorded by accelerometers of type Accelerometer 3). The evaluation is quantitatively carried out through the calculation of the fitting score
\begin{equation}
   \text{fit}=100 \times \frac{1-\norm {a_{v,m}-\hat{a}_v }}{\norm {a_{v,m}-\bar{a}_{v,m}}}
    \label{eq:fit_eq},
\end{equation}
which indicates how well the model prediction fits the validation data sets. In Eq.~\eqref{eq:fit_eq} $a_{v,m}$ is the measured vertical acceleration, $\hat{a}_v$ denotes the acceleration predicted by the identified model, and  $\bar{a}_{v,m}$ is the mean value of the measured acceleration.

To check the validity of the results obtained from the identified models, data sets collected during the receptance tests (called ``validation data sets'') are used. Again, validation data sets are randomly chosen from the measured acceleration responses excluding the data sets used for identification . In each single hammer excitation (or wheel excitation) the initial condition $\mathbf{x}_{0,j}$ is recursively estimated, where $j$ is the impact number (or wheel number). The zero-input output response of the identified model is then obtained using Eq.~\eqref{eq:pulseResponseinitial}.

\begin{figure}[tbp]
	\centering
	\includegraphics[width=0.8\linewidth]{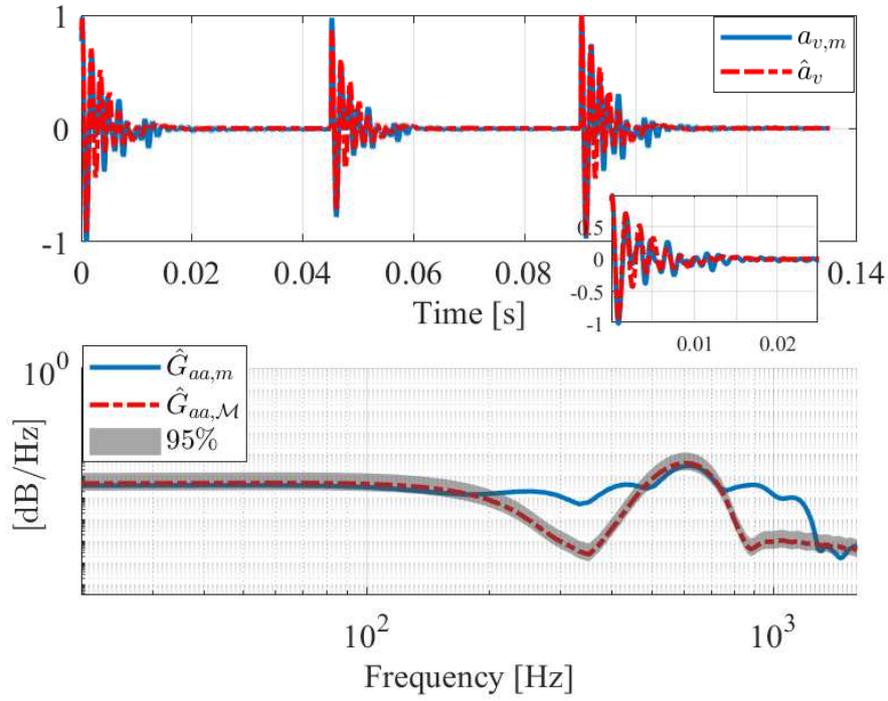}
	\caption{Validation of the identified model on receptance test data at location A7: (top) time domain responses; (bottom) estimated power spectral densities with $95 \%$ confidence interval for the measured data, $\hat{G}_{aa,m}$, and the model, $\hat{G}_{aa,\mathcal{M}}$.}
	\label{fig:validA7}
\end{figure}

Figure~\ref{fig:validA7} compares the validation data set measured at the crossing panel with the predicted response in both time and frequency domains. The identified model well predicts the dominant behavior of the system since the first and second track resonance frequencies in the range $[0,800]\,\mathrm{Hz}$ are matched by the model. The fitting score for the shown validation data set is $73.3\%$.

\begin{figure}[tbp]
	\centering
	\includegraphics[width=0.7\linewidth]{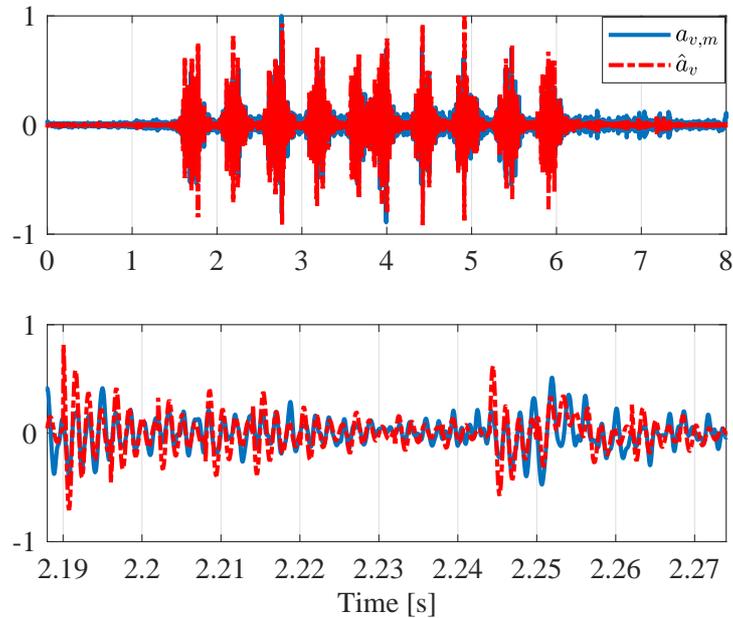}
	\caption{Comparison of measured and predicted accelerations for a IC3 train travelling at 160 km/h through location A7: (top) whole train passage; (bottom) zoom in on a two wheel-set passage.}
	\label{fig:RobustA7}
\end{figure}

To assess the robustness of the model, its capability of predicting the band-pass filtered ($[80,800]\,\mathrm{Hz}$) acceleration response of the system to a train excitation is examined. The predicted acceleration response is compared to the response measured at the turnout position A7, as shown in Fig.~\ref{fig:RobustA7}. The fitting score in this case is 63.9\%. Based on the graphical comparison and the fitting score it is concluded that the identified model for the crossing panel well predicts the dynamic response of the track when excited through a train passage.
Table~\ref{table4} lists the fitting scores of the models identified for the different S\&C sections calculated for both the validation data sets (receptance tests) and data sets collected during train passages both in the main and diverging tracks. To properly evaluate the predictive capability of the identified models against the train passage data, at each location along the main track the fitting score is obtained by averaging the fitting scores computed for a batch of 20 IC3 trains travelling at $160\,\mathrm{km}/\mathrm{h}$. Since less than $1\%$ of the trains go through the diverging track, the fitting score is computed by averaging through 5 IC3 trains travelling at $160\,\mathrm{km}/\mathrm{h}$.
\begin{table}[tbp] \centering \small  
	\caption{Prediction capability of identified models.}\label{table4}
	\begin{tabular}{ l*{6}{c}rl }
	\toprule
	\multirow{2}[3]{*}{Section} & \multicolumn{3}{c}{Fit [\%]} \\ 
	\cmidrule{2-4}
	& Validation & Main track & Diverging track \\
	\midrule
   	OT & $78.3$ & $58.7 \pm 4.1$ & $-$ \\  
	SP & $72.3$ & $53.4 \pm 2.4$ & $-$ \\
	CLP  & $62.3$  & $-$  & $-$ \\
	CP (A7)  & $73.3$ & $63.9 \pm 4.8$ & $-$  \\
	CP (A8)  & $68.2$ & $30.4 \pm 3.5$ & $-$   \\ 
	CP (A9)  & $60.2$ &-& $55.9 \pm 4.3$   \\
	AS (main track)  & $62.3$ & $54.5 \pm 4.4$ & $-$   \\ 
	AS (diverging track) & $67.0$  & $-$ &  $53.4 \pm 4.9$ \\
	\bottomrule
    \end{tabular}
\end{table}
With the exception of the crossing panel location CP(A8), the identified low-complexity behavioral models correctly forecast more than $50\%$ of the S\&C track response induced by train excitation. It is worth noting that the receptance test data recorded at the location CP(A8) is not valid for the frequency range higher than $300 \mathrm{Hz}$, therefore the identified model is just of second order. This explains the $30\%$ fitting score.

\begin{figure}[t] 
	\centering
	\subfloat[Location A1]{\includegraphics[width=0.45\columnwidth]{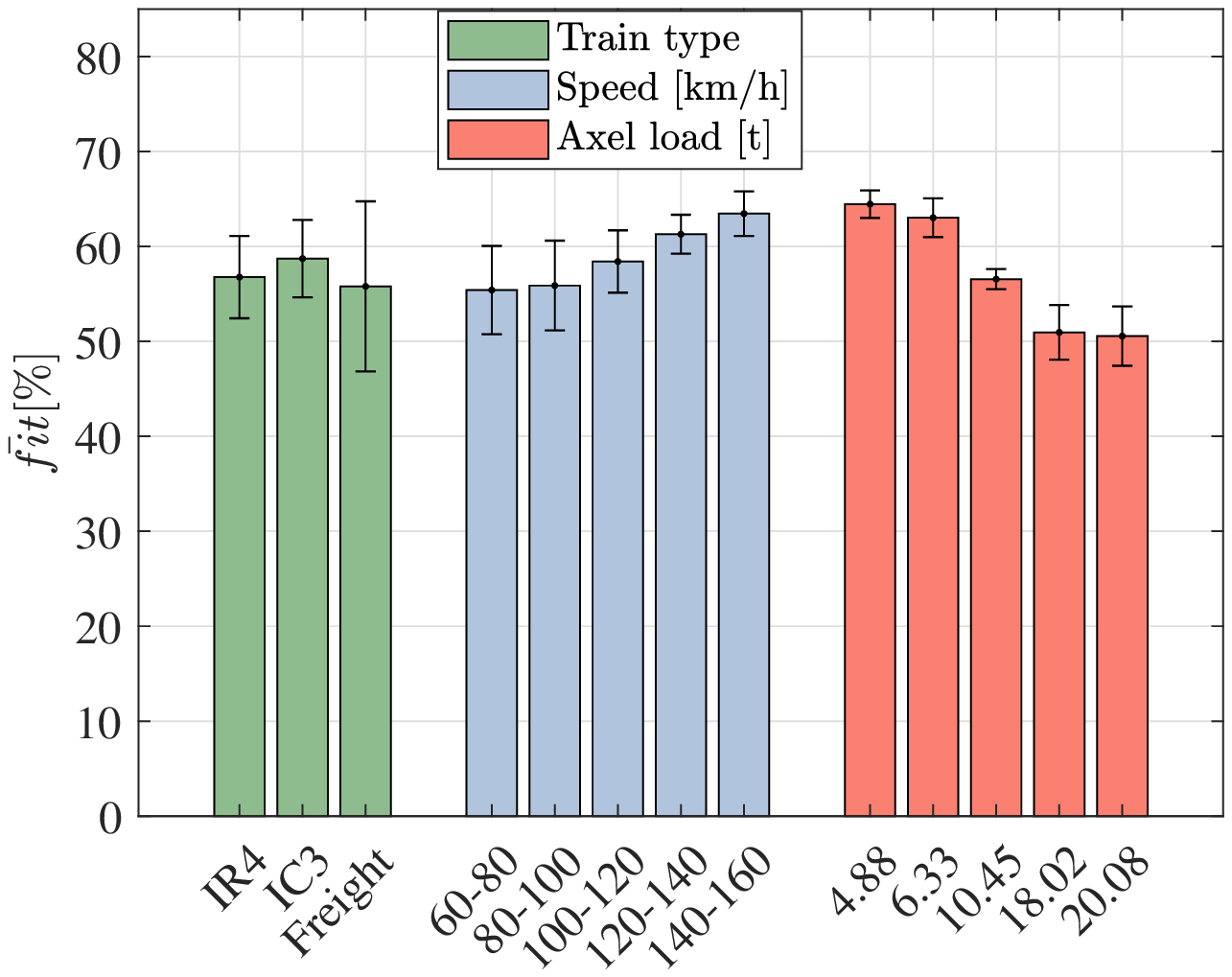}}
	\hspace{0.05cm}	
	\subfloat[Location A2]{\includegraphics[width=0.45\columnwidth]{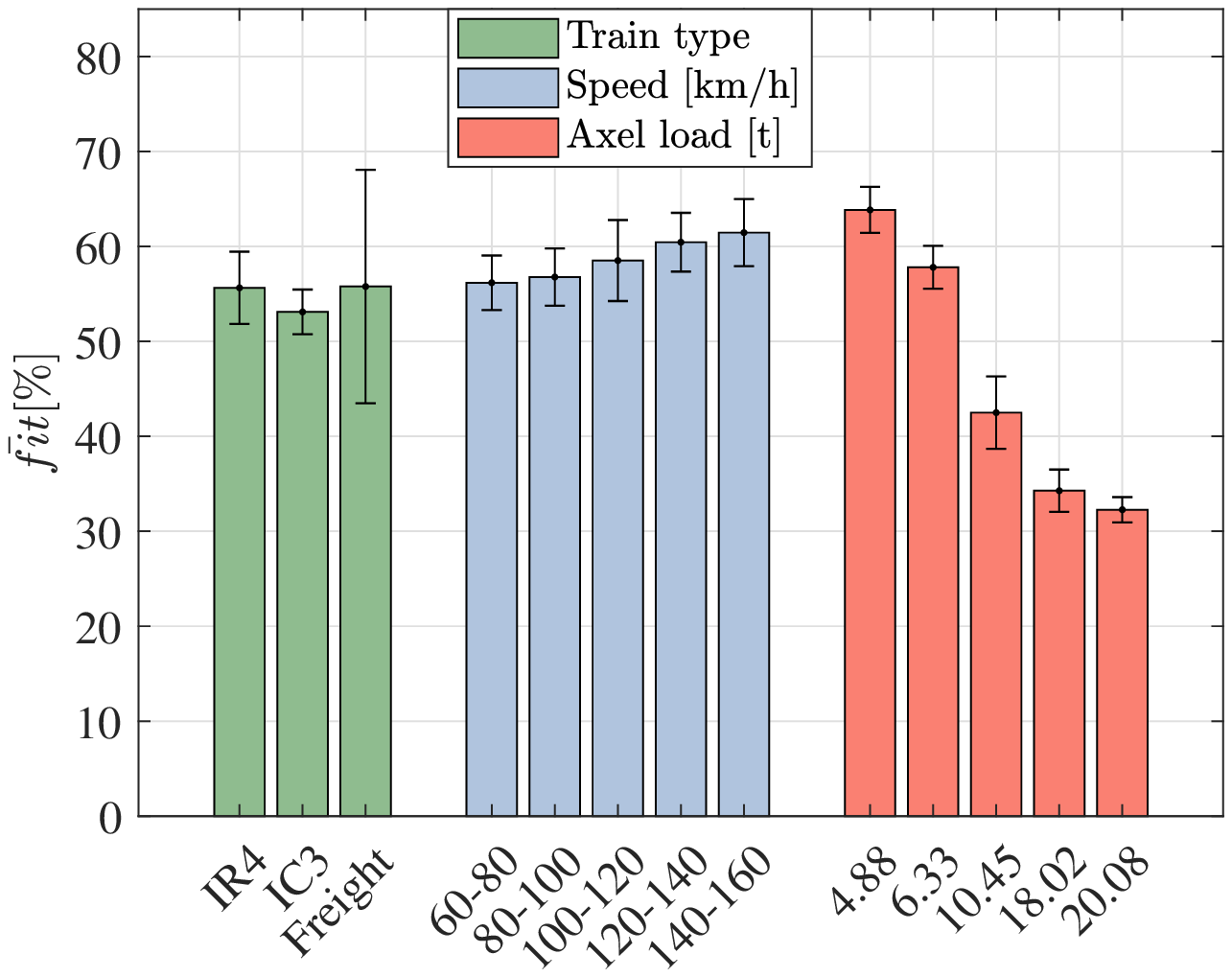}}\\
	\subfloat[Location A7]{\includegraphics[width=0.45\columnwidth]{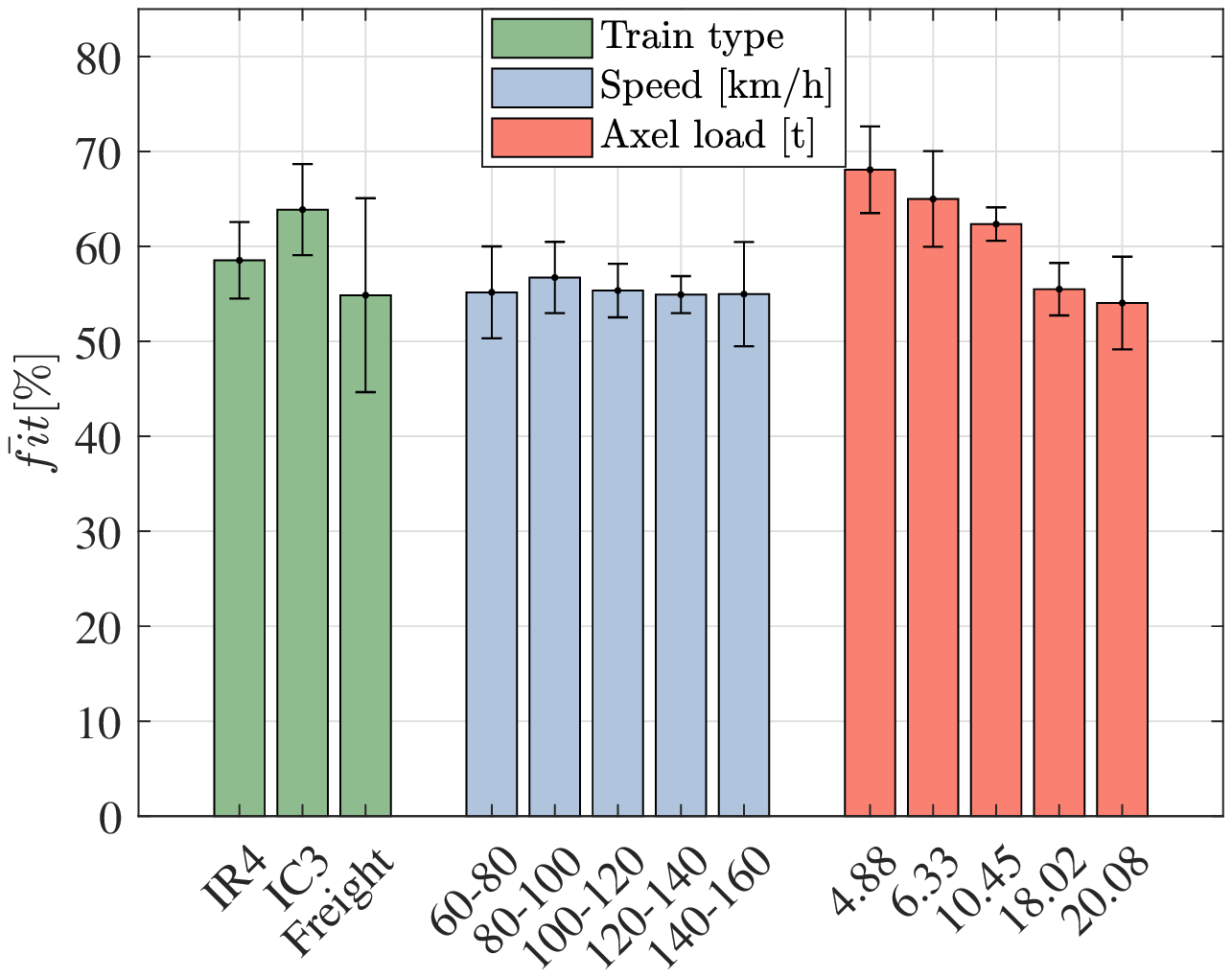}}
	\hspace{0.05cm}	
	\subfloat[Location A11]{\includegraphics[width=0.45\columnwidth]{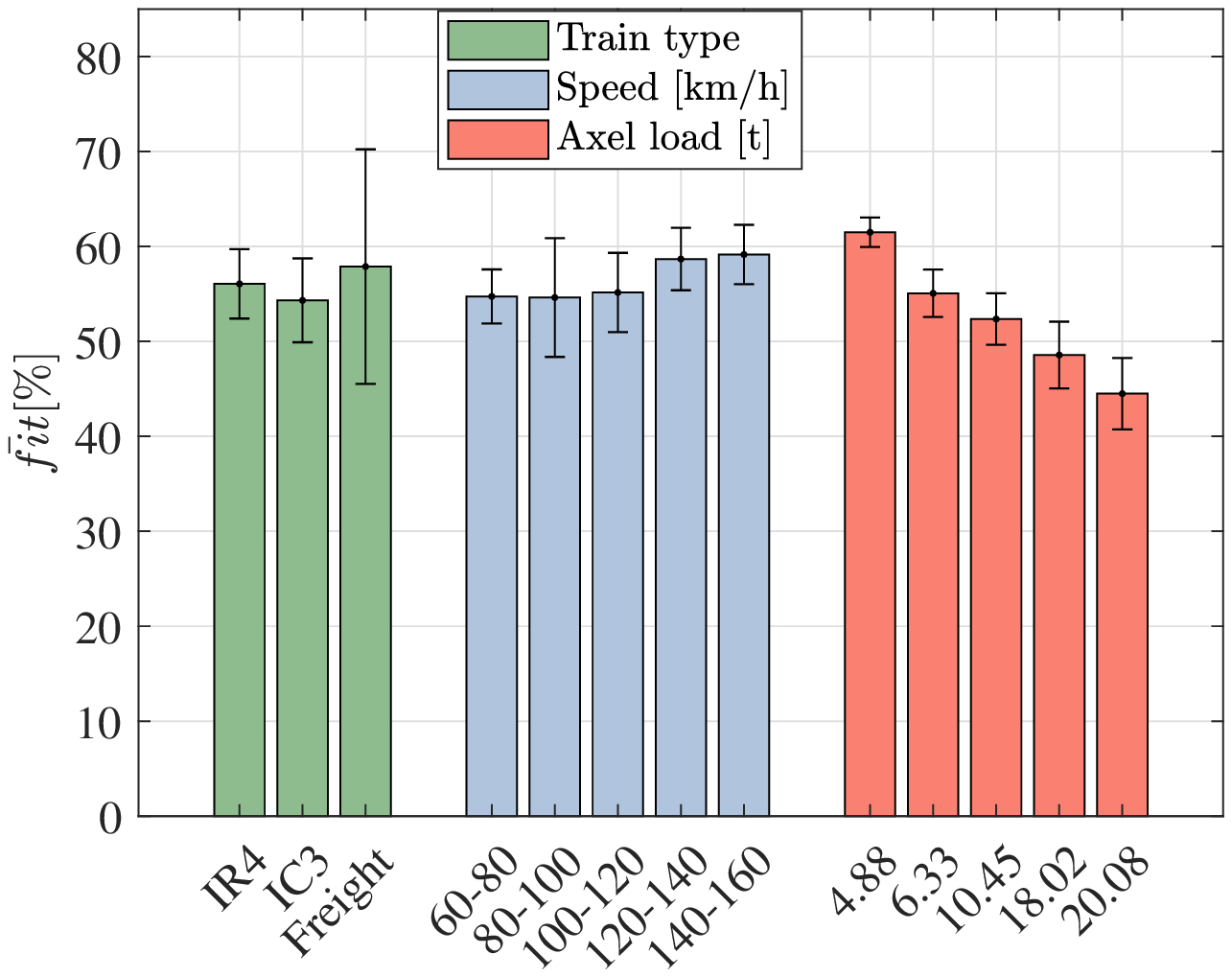}}
	\caption{Average fitting scores and their standard deviation for 100 trains.}
	\label{fig:barcharts}
\end{figure}

The sensitivity and performance of the identified models is investigated in more details to evaluate which parameter among train speed, axle load and train type has the greater influence on the obtained fitting score. For this purpose a batch of 100 trains is considered and the data is clustered in three groups based on train type, train speed and axle load. Figure~\ref{fig:barcharts} illustrates the average fitting scores calculated for different clusters at different sections of the turnout. The predictiveness of the identified models is clearly not dependent on the train type; however trends are visible for the train speed and for the axle load. For all four considered S\&C sections an increase of the axle load results in a reduction of the fitting score from a maximum of about 70\% for a load of $4.88\,\mathrm{t}$ at location A7 to a minimum of about 30\% for a load of $20.08\,\mathrm{t}$ at location A2. The speed dependence is also evident at locations A1, A2 and A11 where the predictiveness increases with train speed. At location A7 the available data do not show any significant trend between the train speed and the model predictiveness.

\section{High-fidelity multi-body simulation model} 
High-fidelity multi-body simulation (MBS) models are theoretical tools that provide a good understanding of the train/track interaction phenomenon, particularly at track sections characterized by complex rail/wheel interface geometries. Similar models have been previously used to evaluate the train/track interaction at railway turnouts  \cite{kassa2006simulation,lagos2012rail,schupp2004modelling,paalsson2015dynamic}. MBS models are appropriate tools to evaluate complex  dynamic interaction problems in the railway field, providing a good balance between the accuracy achieved when simulating the wheel/rail interaction and the computational time.

A particular MBS model developed through the commercial software GENSYS is used to evaluate the dynamic interaction between the train and the track when a passenger train passes through the S\&C. Similar ways of simulating the train/track interaction at railway turnouts by using GENSYS can be found in \cite{kassa2006simulation,kassa2008dynamic}. The high-fidelity MBS model is able to account for the geometric variations of the rail and track substructure along the entire switch and crossing. A predefined number of cross section profiles of the rail are considered to define the geometry of the track model. They actually make up the mesh of the model, so the finer the mesh the more accurate the outputs of the dynamic interaction coming from the MBS model. A similar model was previously defined and used in~\cite{de2017numerical} to properly represent the elastic properties of the track elements in GENSYS.

\begin{figure}[tbp]
	\centering
	\includegraphics[width=\linewidth]{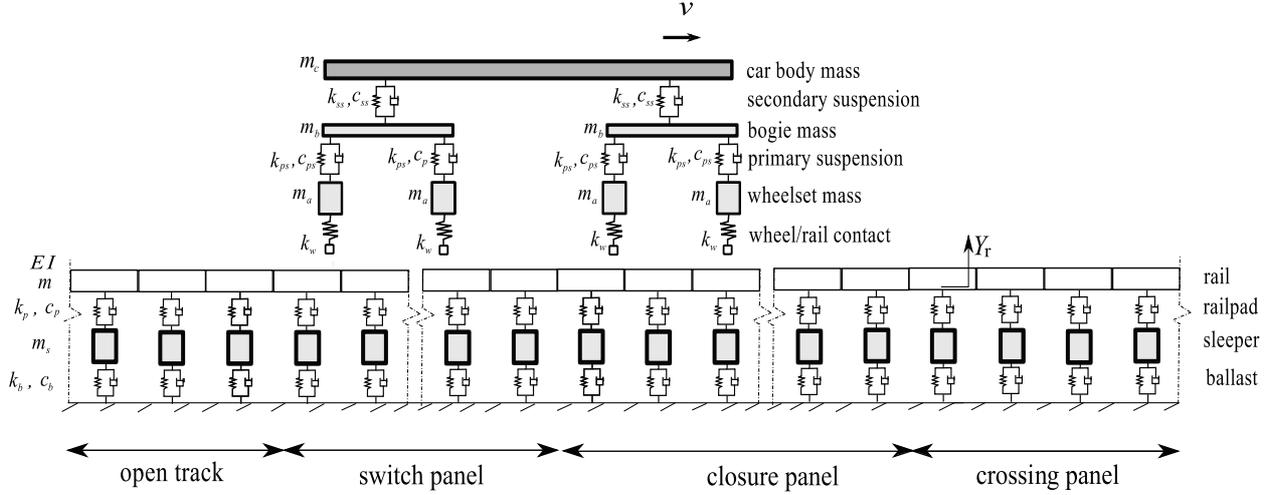}
	\caption{Layout of the high-fidelity MBS model to simulate the train and track interaction.}
	\label{fig:MBSmodel}
\end{figure}

The turnout model consists of a set of mass-spring-damper systems that provide support to a continuous multi-span Euler Bernoulli beam, used to model the rails. The train model consists likewise of a set of mass-spring-damper systems representing the different elements of an IC3 passenger train with a single car body, two bogies and four wheelsets. A layout of the MBS model used to simulate the train and track interaction is depicted in Fig.~\ref{fig:MBSmodel}. Table~\ref{tab:mechanical parameters} provides the mechanical characteristics used to model the main components that constitute the vehicle and the track.

\begin{table}[tbp] \centering
	\caption{Nomenclature of mechanical parameters of Fig.~\ref{fig:MBSmodel}.}\label{tab:mechanical parameters}
	\begin{tabular}{ll} 
	\toprule
	Vehicle & Turnout\\
	\midrule
		 $v$ - train speed & $Y_r$ - vertical displacement of the rail\\
		 $k_{ss}$ - secondary suspension stiffness & $EI$ - bending stiffness of the rail\\
		 $c_{ss}$ - secondary suspension damping & $m$ - mass of the rail\\
		 $m_{b}$ - bogie mass & $k_{p}$ - railpad stiffness\\
		 $k_{ps}$ - primary suspension stiffness & $c_{p}$ - railpad damping\\
		 $c_{ps}$ - primary suspension damping & $m_{s}$ - ballast mass\\
		 $m_{a}$ - wheelset mass & $k_{b}$ - ballast stiffness\\
		 $k_{w}$ - wheel/rail contact & $c_{b}$ - ballast damping\\
		\bottomrule
	\end{tabular}
\end{table}

\subsection{MBS model validation} \label{sec:MBS}
\begin{figure}[tbp]
	\centering
	\subfloat[Location A1]{\includegraphics[width=0.5\textwidth]{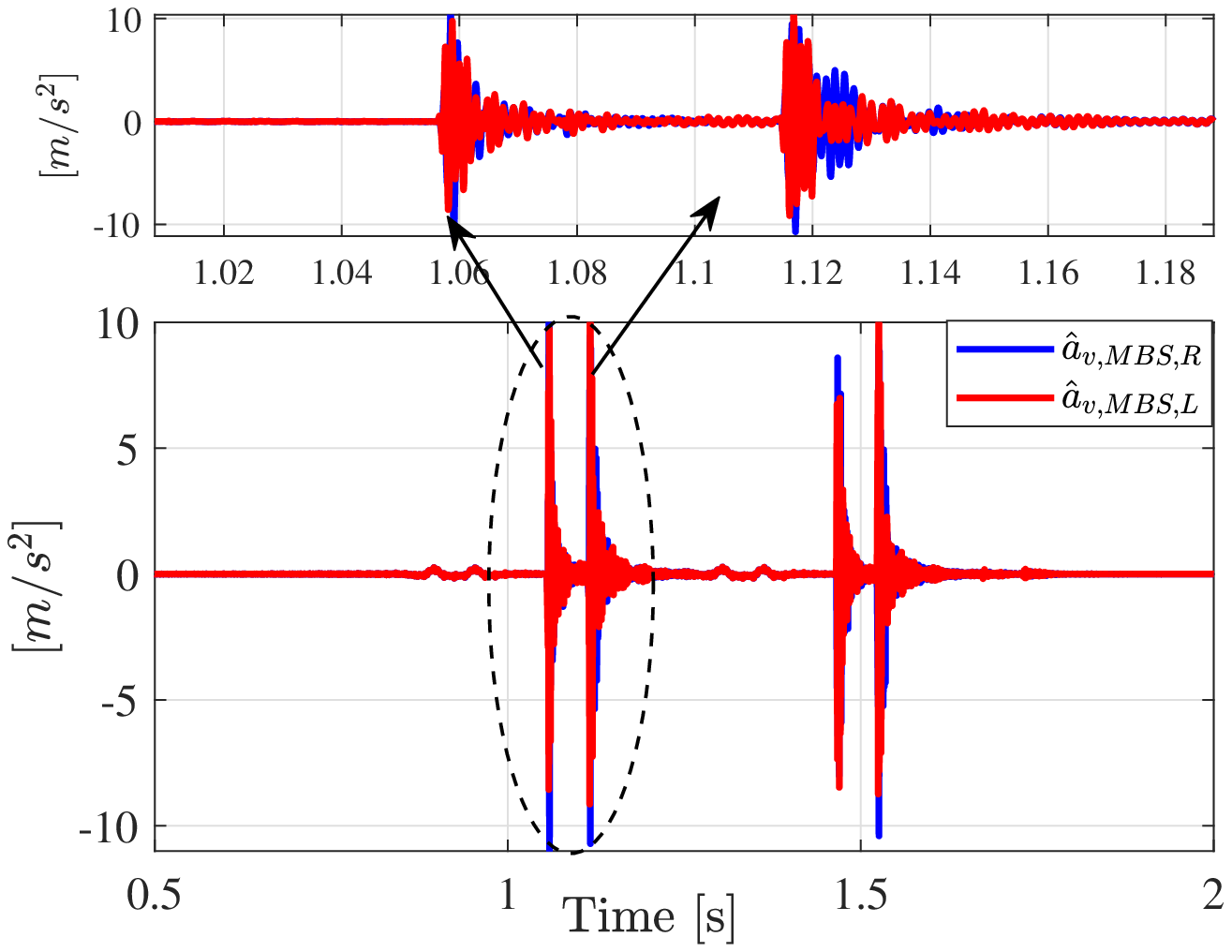}}
	\hfill
	\subfloat[Location A4]{\includegraphics[width=0.5\textwidth]{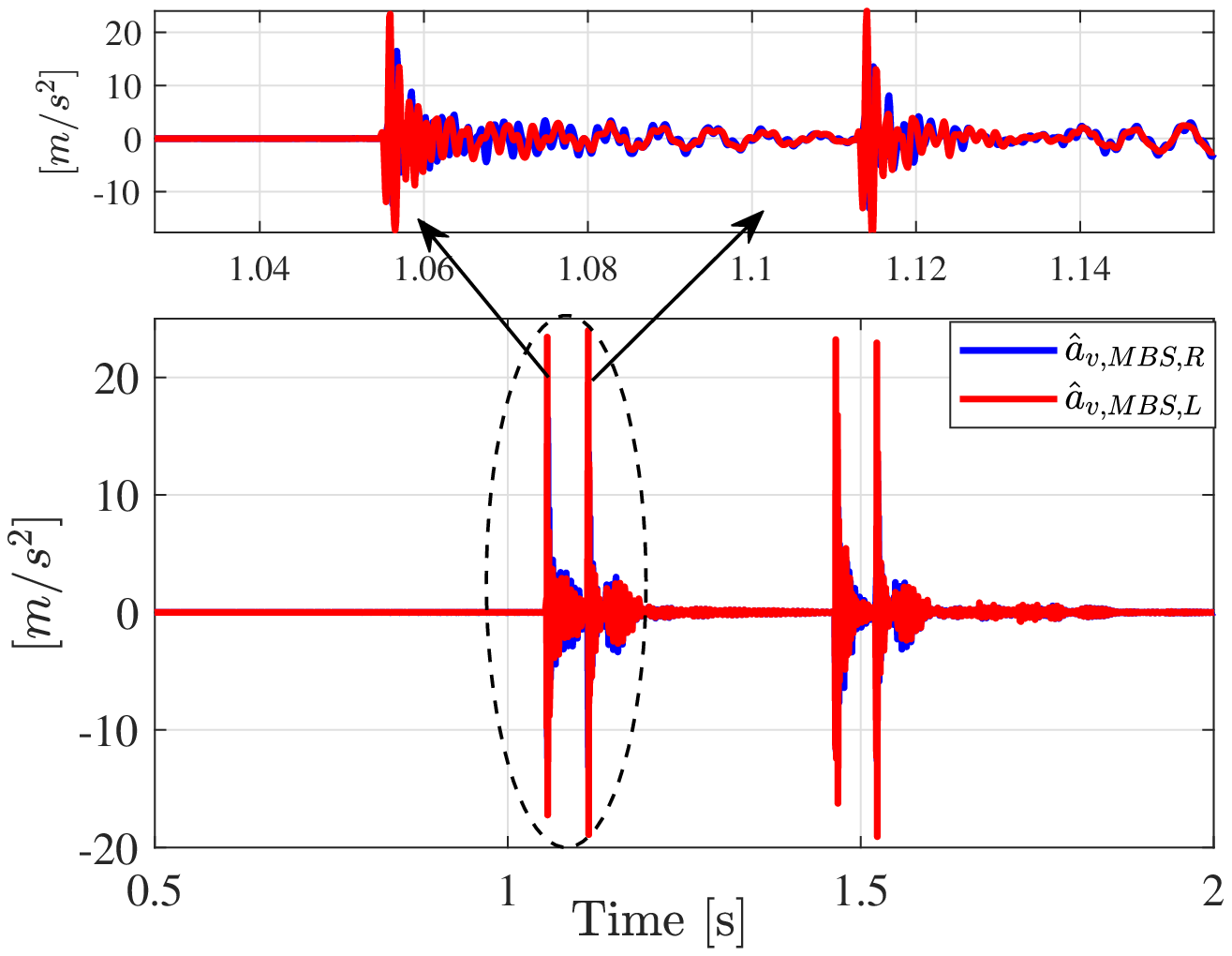}}
	\hfill
	\caption{Vertical accelerations measured on the rail using the high-fidelity MBS model. Top zoomed-in figure passage of first bogie }
	\label{fig:MBSACC}
\end{figure}

The simulation of the dynamic interaction between a single car body and the turnout is carried out in the MBS program. It is considered that for this particular case the train moves through the main track. The time step to perform the numerical simulation is set to $\Delta t=0.0001$ seconds, which allows capturing the most relevant frequencies in the numerical train/track interaction up to $5\,\mathrm{kHz}$. The total computational time to run a single simulation of the train track interaction is 38 minutes. Figure~\ref{fig:MBSACC} illustrates the magnitude of the vertical acceleration obtained with the MBS model, measured on the rail, at locations A1 and A4 along the turnout.

Figure~\ref{fig:MBSACC}(a) shows that the MBS model properly captures the passage of the first bogie at location A1, followed by the impact between the first wheelset and the switch blade on the left side of the switch panel. It is worth highlighting that this impact is commonly detected when analyzing time history responses in MBS models for switches and crossings~\cite{kassa2006simulation,kassa2008dynamic}. The accuracy in terms of magnitude of the acceleration strongly depends on the mesh definition of the MBS code (number of cross sections of the rail profile) at both the switch panel and the crossing panel and it might compromise the computational efficiency of the MBS model if the number of cross sections selected to model the geometric variations along the turnouts is too high. After the impact between the leading wheelset and the switch blade, characterized by a high frequencies content, the passage of the second bogie over the measurement point A1 can be detected, again followed by the impact between the trailing wheelset and the switch blade. In Fig.~\ref{fig:MBSACC}(b) the same phenomenon can be observed, but in this case both the bogie passage patterns and the wheelsets-switch blade impact patterns are nearly overlapped due to the proximity between the measurement point A4 and the location of the switch blade. 

To validate the predictiveness of the MBS model, the simulated vertical acceleration is compared with the vertical acceleration measured on the S\&C at Tommerup station. To emphasize the contribution of different phenomena into the vertical track acceleration the analysis is performed for the low frequency range, $f \in [0,40]\, \mathrm{Hz}$, and for the high-frequency range, $f \in [40,1000]\,\mathrm{Hz}$.

\begin{figure}[tbp]
	\centering
	\subfloat[Time history]{\includegraphics[width=0.5\textwidth]{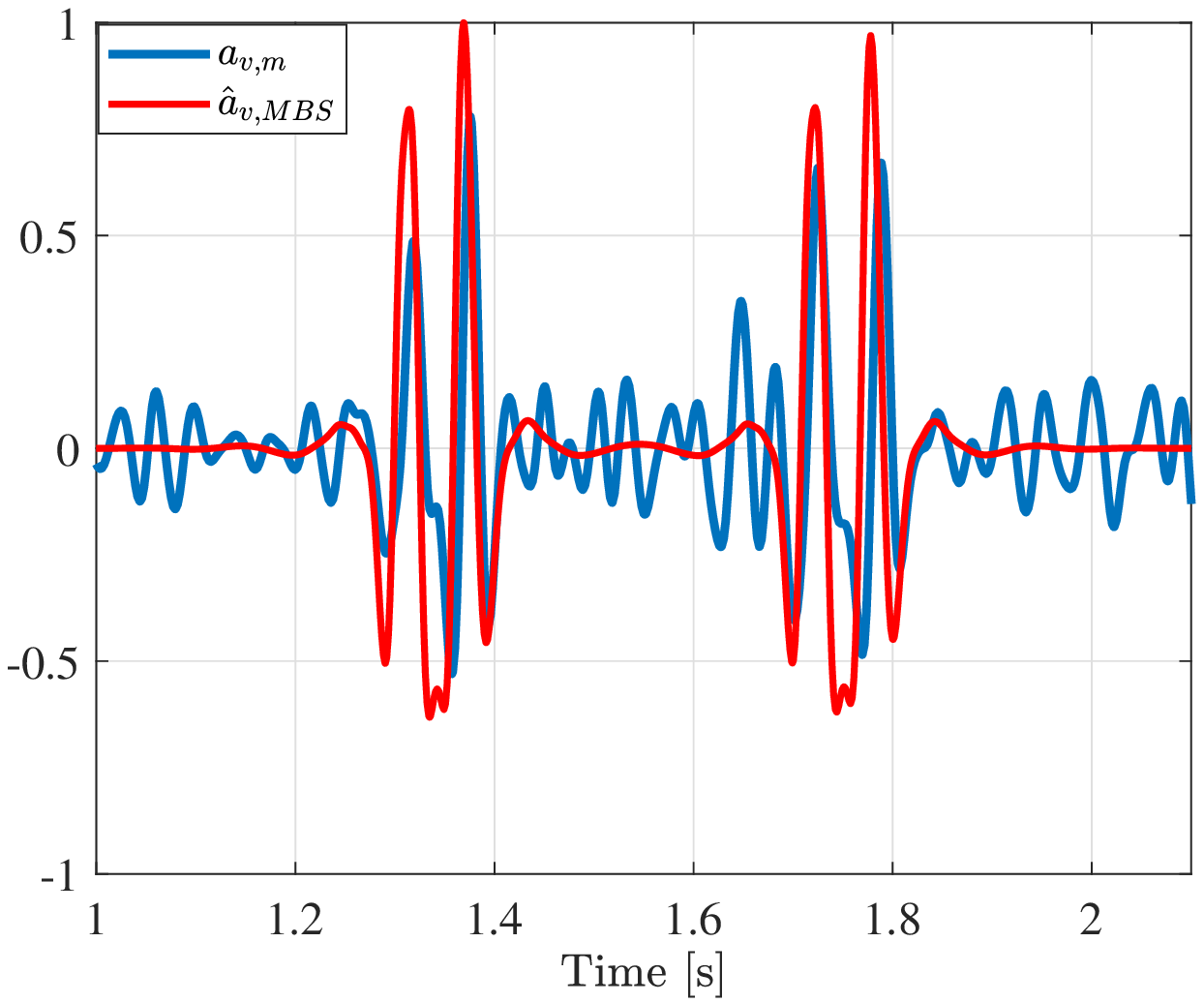}}
	\hfill
	\subfloat[Power spectral density]{\includegraphics[width=0.5\textwidth]{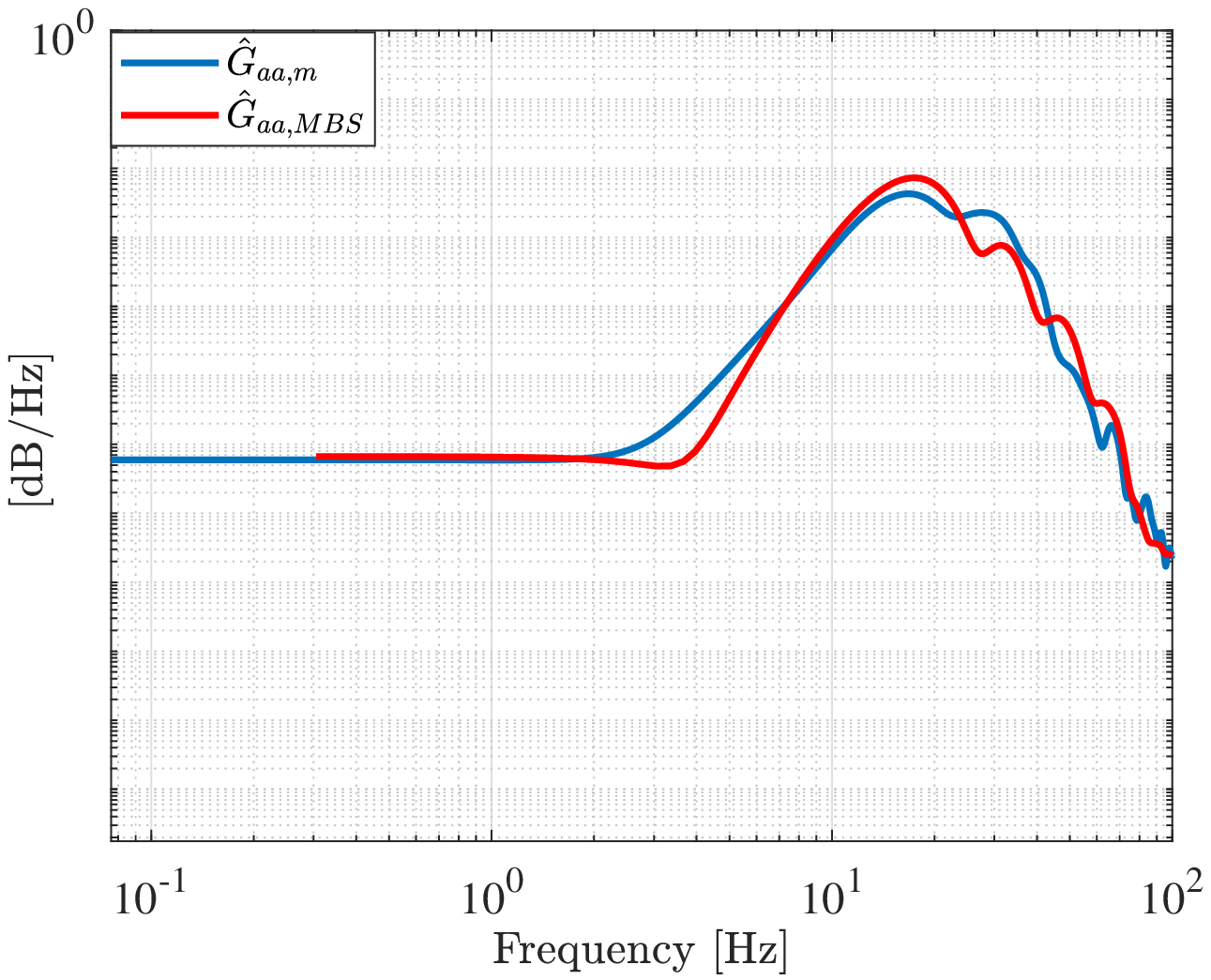}}
	\hfill
	\caption{Measured and simulated vertical accelerations at location A4 -- Low-frequency range.}
	\label{fig:MBSfilter1time}
\end{figure}

\begin{figure}[tbp]
	\centering
	\subfloat[Time history]{\includegraphics[width=0.5\textwidth]{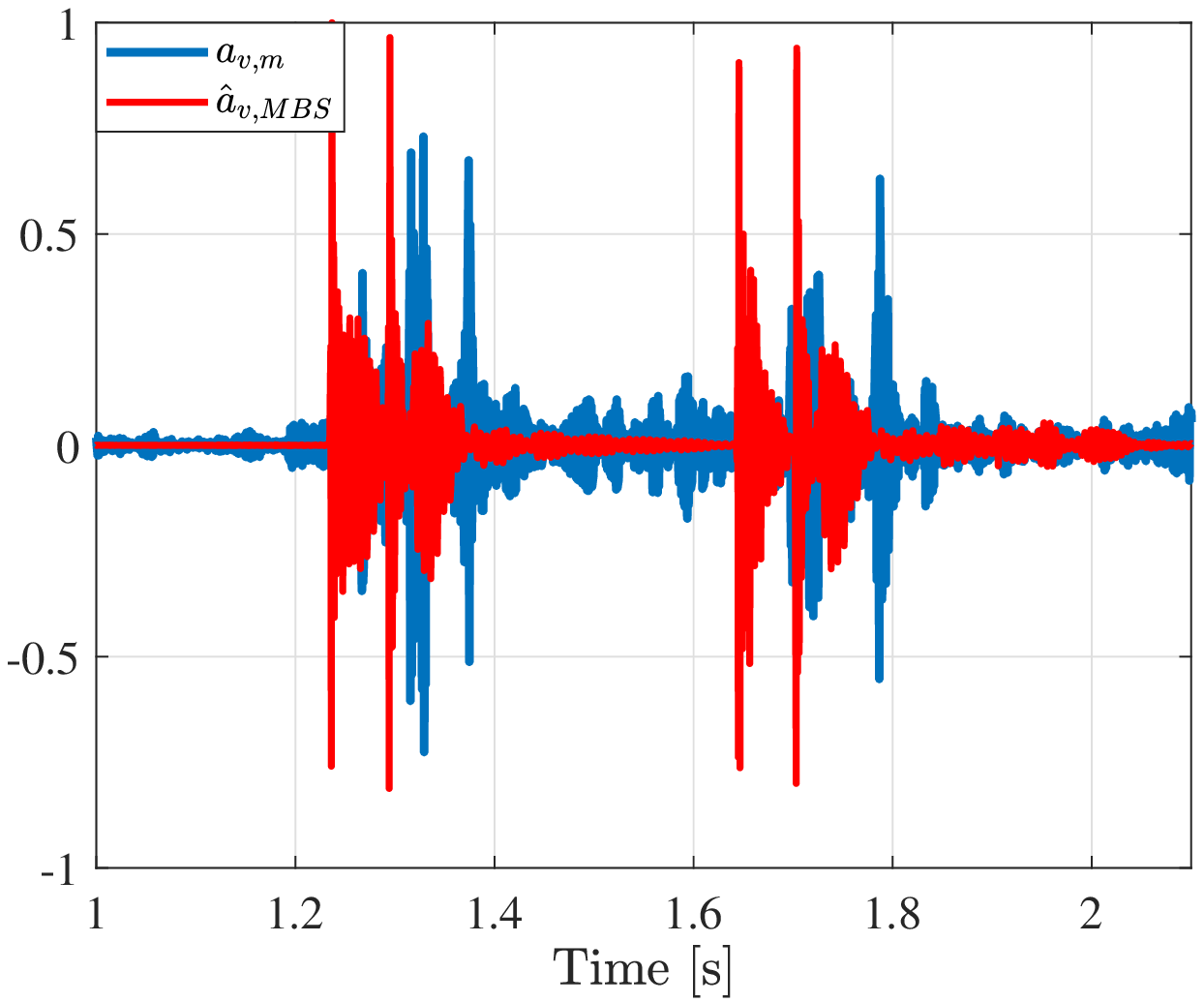}}
	\hfill
	\subfloat[Power spectral density]{\includegraphics[width=0.5\textwidth]{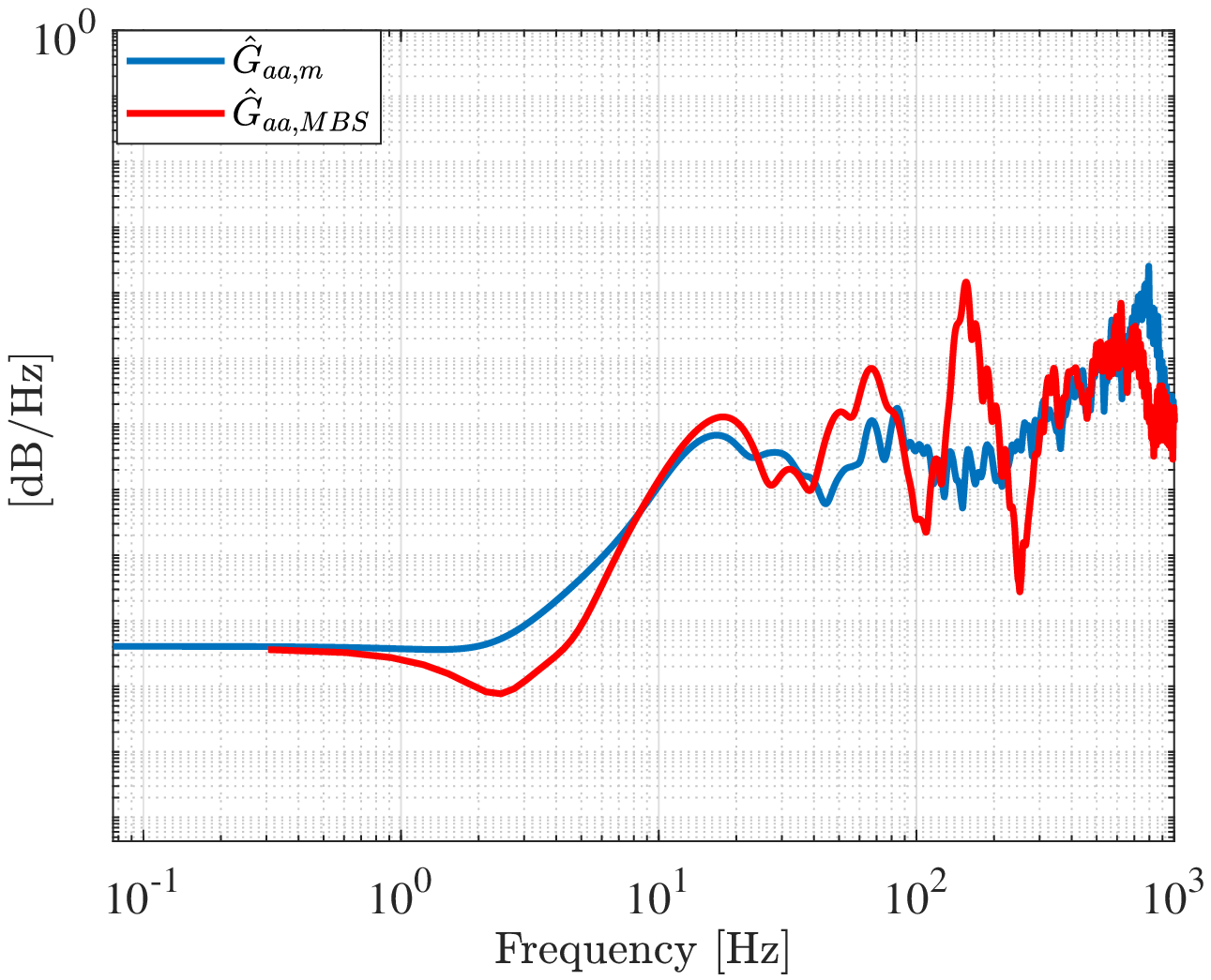}}
	\hfill
	\caption{Measured and simulated vertical accelerations at location A4 -- High-frequency range.}
	\label{fig:MBSfilter2time}
\end{figure}

Figure~\ref{fig:MBSfilter1time} shows the comparison between measured and simulated vertical accelerations in the low-frequency region. Both the time series and the power spectral densities indicate the good ability of the MBS model to correctly reproduce the dynamical behaviour of the train/track interaction and particularly the phenomenon taking place during the passage of bogies through the measurement points. Figure~\ref{fig:MBSfilter2time} shows the comparison between measured and simulated vertical acceleration in the high-frequency region. The MBS model is not capable of predicting measured track response with sufficient accuracy in this frequency range, with a significant overestimation of the power spectral density in $f\in[40,300]\,\mathrm{Hz}$. Although the MBS model was calibrated exploiting the receptance data, its current level of complexity in terms of number of considered rail cross sections does not result in sufficient accuracy to predict the track response in the frequency interval where the ballast and the railpad dynamics are dominant.

\section{Low-complexity behavioural model and high-fidelity MBS model for railway turnouts}
Switches and crossings are a key element of railway networks as they enable efficient and flexible train operations. They are also the most vulnerable part of the railway infrastructure, exposed to higher dynamic loads because of moving parts and the greater geometric complexity compared to normal tracks. 

The continuous development of models to understand how the different track components contribute to the dynamical behaviour of the S\&C upon train passage is of paramount importance to improve the know-how needed for advancing the turnout technology. The ultimate purpose of the model determines which features are fundamental, thereby influencing the choice of modelling techniques and tools.

For the purpose of investigating train/track interaction features as predictiveness, accuracy and precision are preeminent. However high accuracy and precision generally come at the cost of large complexity that, in turn, determines a high computational burden. Therefore, to limit the intricacy, models developed for studying the train/track interaction usually aim at achieving very good predictions of track response due to the wheel-rail contact forces, thereby focusing on the low-frequency range ($f < 100\,\mathrm{Hz}$) of the track response. To obtain high predictiveness also in relation to the track resonances (ballast, railpad and rail), the model complexity should be further increased. Hence this type of models are difficult to scale and to adapt to different type of S\&Cs. The adopted MBS model clearly highlights the strengths and weaknesses of this modeling approach.

For the purpose of monitoring the health state of the S\&C features as portability, robustness, scalability and predictiveness are instead deemed paramount. A modelling tool embodying these characteristics will secure deployment across the entire railway network thanks to the inherent ability to adapt to changes in S\&C technology, operational conditions and environmental settings.

The developed low-complexity behavioural model encompasses the aforementioned features. The model can be adapted to a different S\&C by using new receptance data, making it portable across different network locations. If an additional track component needs to be monitored, then the model can be easily expanded by increasing the model order; e.g., the dynamics linked to the rail resonance frequency can be included by augmenting the model order from four to six. The predictive power and the robustness have been demonstrated by testing the model on data sets differing for train travelling speed and axle load (see Fig.~\ref{fig:barcharts} and Table~\ref{table4}).

The capability of the proposed model to reliably estimate the dynamic behavior of the turnout related to the ballast layer and railpads opens opportunities for the determination of deviations from the current level of wear and tear. In fact, degradation processes occurring in the ballast and railpad result in changes of the stiffness of these track infrastructure components which, in turn, influence the first and second track resonance frequencies \cite{lam2012feasibility,oregui2017sensitivity}. Therefore, long-term monitoring of the quality of the track components can be performed through the recursive estimation of the track resonance frequencies and damping ratios over time, using the identified low-complexity behavioral model.

\section{Conclusions}
The paper proposed a new modelling approach for railway turnouts that based on subspace system identification techniques provides a low-complexity behavioural model that describes the dominant dynamics related to the first two track resonances associated with the ballast and railpad.

The modeling approach exploits vertical track accelerations measured on the rail head during a receptance test campaign in combination with the Eigensystem Realization Algorithm to identify a fourth order linear model. The identified model is characterized by two resonance frequencies, one related to the ballast layer and one related to the railpad.

Low-complexity behavioural models have been identified for five different sections of an S\&C located at Tommerup station (Fyn, Denmark). The identified resonance frequencies for the different sections are in line with values previously reported in the literature. The identified models have been extensively validated using vertical track acceleration data collected both during the receptance test campaign and train passages. In particular a pool of data related to hundred train passages differing for train type, traveling speed and axle load has been exploited to verify the robustness of the models predictions. The robustness analysis demonstrated that the low-complexity model prediction accuracy increases for high-speed trains with the lowest axle load. For trains traveling with a speed in the range $140 - 160 \, \mathrm{km}/\mathrm{h}$ and with an axle load of $4.88\,\mathrm{t}$ the identified models have a fitting score to the measured data in the interval $60\% - 70\%$.

The proposed modelling approach generates models that feature good predictiveness, robustness to changes in operational conditions, portability throughout the S\&C and across S\&Cs, scalability. These qualities render the low-complexity behavioural models suitable for condition monitoring of the S\&Cs since the development of degradation phenomena affecting track components as the ballast layer and the railpad could be assessed through significant variations in the model parameters over relatively large periods of time.

\section*{Acknowledgements}
This research study has been carried out, as part of the \href{http://www.intelliswitch.dk/}{INTELLISWITCH} project. The research is financially supported by Innovation Fund Denmark with grant number 4109-00003B. The authors gratefully acknowledge this support. 




 \bibliographystyle{elsarticle-num-names}
\bibliography{sample}

\end{document}